\documentclass[10pt,journal,compsoc]{IEEEtran}
\usepackage{amsmath,amssymb,amsfonts}
\usepackage{algorithmic}
 \usepackage[ruled,vlined]{algorithm2e}
\usepackage{graphicx}
\usepackage{textcomp}
\usepackage{xcolor}
\usepackage[hyphens]{url}
\usepackage{hyperref}
\hypersetup{
    colorlinks=true,
    linkcolor=blue,
    filecolor=magenta,      
    urlcolor=cyan
}

\newcommand{\ignore}[1]{}
\usepackage[pass]{geometry}

\usepackage{color}
\usepackage{soul}
\usepackage{mathtools}
\DeclarePairedDelimiter\ceil{\lceil}{\rceil}
\usepackage{booktabs}
\usepackage{subcaption}
\usepackage{subfiles}
\usepackage{verbatim}
\ifCLASSOPTIONcompsoc
  \usepackage[nocompress]{cite}
\else
  \usepackage{cite}
\fi

\ifCLASSINFOpdf
\else
\fi

\begin{document}
\title{High Throughput Multidimensional Tridiagonal Systems Solvers on FPGAs}

\author{Kamalavasan Kamalakkannan, %~\IEEEmembership{Member,~IEEE,}
        Istvan Z. Reguly,%~\IEEEmembership{Member,~IEEE,} \\
        Suhaib A. Fahmy %~\IEEEmembership{Senior Member,~IEEE,} 
        and 
        Gihan R. Mudalige%~\IEEEmembership{Member,~ACM}% <-this % stops a space
\IEEEcompsocitemizethanks{\IEEEcompsocthanksitem Kamalavasan Kamalakkannan 
and Gihan R. Mudalige are with the Department of Computer Science, University 
of Warwick, UK.\protect\\
E-mail: \{kamalavasan.kamalakkannan, g.mudalige\}@warwick.ac.uk
\IEEEcompsocthanksitem Istvan Z. Reguly is with the Faculty of Information 
Technology \& Bionics \\ Pazmany Peter Catholic University, Hungary\protect\\
E-mail: reguly.istvan@itk.ppke.hu
\IEEEcompsocthanksitem Suhaib A. Fahmy is the with the King Abdullah University 
of\\ Science and Technology (KAUST), Saudi Arabia\protect\\
E-mail: suhaib.fahmy@kaust.edu.sa
}% % stops an unwanted space
% \thanks{Manuscript received Oct 27, 2021.}
}

\IEEEtitleabstractindextext{%
\begin{abstract}
This paper presents a design space exploration for synthesizing optimized, 
high-throughput implementations of multiple multi-dimensional tridiagonal 
system solvers on FPGAs. Re-evaluating the characteristics of algorithms for the 
direct solution of tridiagonal systems, we develop a new tridiagonal solver 
library aimed at implementing high-performance computing applications on Xilinx 
FPGA hardware. Key new features of the library are (1) the unification of 
standard state-of-the-art techniques for implementing implicit numerical solvers 
with a number of novel high-gain optimizations such as vectorization and 
batching, motivated by multiple multi-dimensional systems common in real-world 
applications, (2) data-flow techniques that provide application specific 
optimizations for both 2D and 3D problems, including integration of explicit 
loops commonplace in real workloads, and (3) the development of a predictive 
analytic model to explore the design space, and obtain rapid resource and 
performance estimates. The new library provide an order of magnitude better 
performance when solving large batches of systems compared to Xilinx's current 
tridiagonal solver library. Two representative applications are implemented 
using the new solver on a Xilinx Alveo U280 FPGA, demonstrating over 85\% 
predictive model accuracy. These are compared with a current state-of-the-art 
GPU library for solving multi-dimensional tridiagonal systems on an Nvidia V100 
GPU, analyzing time to solution, bandwidth, and energy consumption. Results show 
the FPGAs achieving competitive or better runtime performance for a range of 
multi-dimensional mesh problems compared to the V100 GPU. Additionally, the 
significant energy savings offered by FPGA implementations, over 30\% for the 
most complex application, are quantified. We discuss the algorithmic trade-offs 
required to obtain good performance on FPGAs, giving insights into the 
feasibility and profitability of FPGA implementations.
\end{abstract}

\begin{IEEEkeywords}
Multidimensional tridiagonal solvers, high level synthesis, field programmable gate arrays.
\end{IEEEkeywords}}

\maketitle

\IEEEdisplaynontitleabstractindextext

% \fi
\IEEEpeerreviewmaketitle

\IEEEraisesectionheading{\section{Introduction}\label{sec:introduction}}

%%% Contents start here
\IEEEPARstart{T}{ridiagonal}  systems of equations are solved in a wide range of 
High Performance Computing (HPC) applications, particularly as part of the 
numerical approximation of multi-dimensional partial differential equations 
(PDEs). In computational finance, the frequently used Alternating Direction 
Implicit (ADI) time discretization (see Paceman and 
Rachford~\cite{PeacemanAndRachford1955}, and Douglas and 
Gunn~\cite{DouglasAndGunn1964}) leads to the need to solve multiple tridiagonal 
systems of equations in each dimension. In computational fluid dynamics (CFD), 
tridiagonal systems form the core component for using implicit 
techniques~\cite{Pulliam1986} with applications in solving incompressible fluid 
flow problems~\cite{Wang2013} and design of turbo-machinery~\cite{Brandvik2011}, 
among others. The large number of independent tridiagonal systems, often in 
multiple dimensions, offer significant parallelization opportunities on modern 
multi-core and many-core architectures. Recent works such as L\'{a}szl\'{o} et 
al.~\cite{Laszlo2016} demonstrated significant speedups, re-evaluating the well 
known tridiagonal solver algorithms, Thomas~\cite{thomas1949}, 
PCR~\cite{GanderGolub1997}, and their combinations. 

In this paper we evaluate the parallelization opportunities afforded by 
tridiagonal systems solver algorithms on modern FPGA hardware devices. The 
data-flow parallelism targeted in an FPGA provides significant scope to exploit 
the parallelism inherent in tridiagonal solvers. As such, our underlying goal is 
to understand the criteria for a given system solver to be amenable to FPGA 
implementation and uncover the limitations and profitability of such 
accelerators. Previous work on tridiagonal system solvers for FPGAs utilized 
both low-level hardware description languages~\cite{ARC2008, ICCM2012, 
ReconfigTS2012} as well as high-level synthesis tools~\cite{ICCS2014, PCTAC2014, 
ISCAS2015, RC2019, xilinxlib}. They demonstrated implementation of standard 
tridiagonal system solver algorithms (Thomas, PCR, and Spike), evaluating how to 
best utilize FPGA resources to maximize performance. However, many of these 
previous works only develop single system solvers in isolation without a design 
strategy that can be applied for multiple systems and multi-dimensions in 
general and do not utilize higher-gain optimizations for real-world 
applications. Some apply application specific optimizations which are not 
developed as general synthesis techniques. Comparison of performance to 
traditional architectures such as GPUs are also limited in current literature, 
minimizing insights into the utility of FPGAs for this class of applications. A 
key gap is the lack of a unifying design strategy particularly focusing on 
realistic, non-trivial applications. 

In this paper we attempt to bridge this gap with a unifying workflow for 
designing near-optimal FPGA implementations for these implicit numerical solvers, 
applied to the solution of real-world multi-dimensional applications. More 
specifically we make the following contributions:\vspace{-0.1cm}
\begin{itemize}
\item We consider the standard tridiagonal solver algorithms together with 
state-of-the-art FPGA implementations and re-examine the algorithmic trade-offs 
required for obtaining optimized, high-throughput solutions for multiple solves 
in multiple dimensions. We propose a design and optimization strategy for 
developing FPGA implementations selecting the best designs, based on problem 
size, dimensionality, number of systems solved and data-flow paths required, 
including the utilization of High Bandwidth Memory (HBM) on modern devices for 
combining multiple dimension solves and explicit loops in applications. A key 
optimization, novel in this area is the batched execution of multiple 
independent solves on FPGAs leading to superior performance compared to the 
state-of-the-art, the current Xilinx tridiagonal solver library. 

\item Targeting current generation Xilinx FPGAs we implement our designs to 
produce a new tridiagonal solver library that can be used in the solution of 
multi-dimensional applications. Using this, we present the optimized design of 
two non-trivial applications, a 2D and 3D ADI heat diffusion solve, 
implemented with both single precision (FP32) and double precision (FP64) 
floating point representations, and a 2D Stochastic-Local Volatility (SLV) 
model application from the financial computing domain. Given hardware resource 
constraints, we focus on features of the applications that are amenable for FPGA 
implementation and optimizations for gaining near-optimal, high throughput 
performance.  
% The use of HMB on modern FPGAs is first shown here -- shold note this. 

\item We develop a predictive analytic model that provides estimates for 
application runtime giving insights into the profitability of implementing the 
tridiagonal system solvers on Xilinx FPGAs using our design strategy. The model 
predicts the runtime performance considering system/batch sizes and 
optimizations implemented together with memory requirements and operating 
frequency. Runtime predictions from the model are shown to be within 15\% of the 
achieved runtime on evaluated applications. 

\item Finally, the runtime, bandwidth, and energy performance of the FPGA 
implementations on a Xilinx Alveo U280 are compared with a state-of-the-art 
multi-dimensional tridiagonal solver library for GPUs on the HPC-grade Nvidia 
V100 GPU. 
\end{itemize}
Results on the U280 FPGA demonstrate competitive performance compared to the 
best performance achieved for the same application on the GPU using both FP32 
and FP64 representations. To our knowledge the extended work-flow, new library, 
predictive model and the superior performance demonstrated for the above 
applications in this paper present key innovations, advancing the 
state-of-the-art. We believe our design path provides a promising strategy for 
use with industrial workloads, particularly from the financial computing domain, 
significantly reducing the complexity of the development cycle for these 
platforms. 

The rest of the paper is organized as follows: Section~\ref{sec/background} 
presents a brief overview of tridiagonal solver algorithms together with 
previous work on synthesizing tridiagonal solvers on FPGAs, including the 
current state-of-the-art. Section~\ref{sec/design} presents our 
proposed design strategy, as a step-by-step methodology, starting from the 
basic algorithms, down to target FPGA code for the Xilinx Alveo FPGAs. A 
performance analysis and benchmarking of the FPGA implementations compared 
to the GPU performance is presented in Section~\ref{sec/perf}. Finally, 
conclusions are presented in Section~\ref{sec/conclusions}.\\

\vspace{-0.4cm}
\section{Background}\label{sec/background}

\noindent Tridiagonal systems arise from the need to solve a system of linear 
equations as given in equation (\ref{eq/tdma-single}), where \(a_{0} = c_{N-1} = 
0\). Its matrix form $Ax = d$ can be stated as in equation 
(\ref{eq/tdma-system}). 
\begin{equation}
a_{i}u_{i-1}+b_{i}u_{i}+c_{i}u_{i+1} = d,\ \ i = 0, 1, ..., N - 1
\label{eq/tdma-single}
\end{equation}
\begin{equation}
\small\begin{bmatrix}
b_{0} & c_{0}  & 0      &\dots & 0 \\
a_{1} & b_{1}  & c_{1}  &\dots & 0 \\
0     & a_{2}  & b_{2}  &\dots & 0 \\
\vdots& \vdots & \vdots &\ddots & \vdots \\
0     & 0      & \dots  & a_{N-1} & b_{N-1} 
\end{bmatrix}
\begin{bmatrix}
u_{0} \\
u_{1} \\
u_{2} \\
\vdots \\
u_{N-1}
\end{bmatrix}\hspace{-5pt}
=\hspace{-5pt}
\begin{bmatrix}
d_{0} \\
d_{1} \\
d_{2} \\
\vdots \\
d_{N-1}
\end{bmatrix}
\label{eq/tdma-system}\normalsize
\end{equation}
The solution to such systems of equations are well known. The Thomas 
algorithm~\cite{thomas1949} carries out a specialized form of Gaussian 
elimination providing the least computationally expensive solution, but suffers 
from a loop carried dependency (see Algo.~\ref{alg/thomas}). It has a time 
complexity of $O(N)$.\vspace{-0.3cm}
\begin{algorithm}
% \begin{adjustwidth}
\begin{algorithmic}[1]
\STATE $d_{0}^{*} \gets d_{0} / b_{0}$
\STATE $c_{0}^{*} \gets c_{0} / b_{0}$
\FOR {$i = 1, 2, ..., N-1$}
  \STATE $r \gets 1 / (b_{i} - a_{i}c_{i-1}^{*})$
  \STATE $d_{i}^{*} \gets r (d_{i} - a_{i} d_{i-1}^{*})$
  \STATE $c_{i}^{*} \gets r c_{i}$
\ENDFOR
\FOR {$i = N-2, ..., 1, 0$}
  \STATE $d_{i} \gets d_{i}^{*} - c_{i}^{*}d_{i+1}$
\ENDFOR
\RETURN $d$
\end{algorithmic}
\caption{\texttt{thomas}$(a,b,c,d)$}
\label{alg/thomas}
% \end{adjustwidth}
\end{algorithm}\vspace{-0.3cm}

\begin{figure}[t]
\centering
\includegraphics[width=6.5cm]{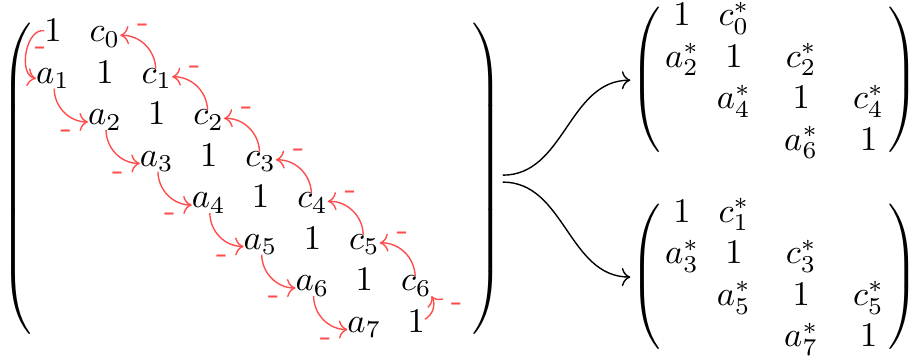} \vspace{-0.2cm}
\caption{One iteration of the PCR algorithm. After the 
iteration every second row will form a separate tridiagonal system.}
\label{fig/PCR-Step}\vspace{-0.5cm}
\end{figure}

In contrast, the PCR algorithm~\cite{GanderGolub1997}~(Algo.~\ref{alg/pcr}), 
operates on a normalized matrix so that \(b_{i} = 1\) and then for each matrix 
row \(i\), subtracts multiples of rows \(i \pm 2^{0}, 2^{1}, 2^{2}, ..., 
2^{P-1}\), where \(P\) is the smallest integer such that \(2^{P} \ge N\). 
\vspace{-0.2cm}
\begin{algorithm}
\begin{algorithmic}[1]
\FOR {$p = 1, 2, ..., P$}
\STATE $s \gets 2^{p-1}$
\FOR {$i = 0, 1, ..., N-1$}
\STATE $r \gets 1 / (1 - a_{i}^{(p-1)} c_{i-s}^{(p-1)} - c_{i}^{(p-1)} 
a_{i+s}^{(p-1)})$
\STATE $a_{i}^{(p)} \gets -r (a_{i}^{(p-1)} a_{i-s}^{(p-1)})$
\STATE $c_{i}^{(p)} \gets -r (c_{i}^{(p-1)} c_{i+s}^{(p-1)})$
\STATE $d_{i}^{(p)} \gets r (d_{i}^{(p-1)} - a_{i}^{(p-1)} d_{i-s}^{(p-1)} - 
c_{i}^{(p-1)} d_{i+s}^{(p-1)})$
\ENDFOR
\ENDFOR
\RETURN $d^{(P)}$
\end{algorithmic}
\caption{\texttt{pcr}$(a,b,c,d)$}
\label{alg/pcr}
\end{algorithm}\vspace{-0.2cm}
This leads to each iteration reducing each of the current systems into two 
systems of half the size (see ~Fig~\ref{fig/PCR-Step}). After \(P\) steps, all 
of the modified \(a\) and \(c\) coefficients are zero, leaving values for the 
unknowns \(u_{i}\). In PCR, the iterations of the inner loop do not depend on 
each other, which is well suited for traditional multi-core/many-core 
architectures such as CPUs and GPUs allowing multiple threads to be used to 
solve each tridiagonal system. However, PCR has a complexity of $O(N \log N)$ 
and is more computationally expensive than the Thomas algorithm, which for an 
FPGA implementation poses an important consideration, (which we will examine in 
Section~\ref{sec/design}) due to the limited availability of resources.

The Spike algorithm~\cite{Polizzi2006} decomposes the $A$ matrix, into $p$ 
partitions of size $m$ to obtain the factorization of $A = DS$ where $D$ is a 
main diagonal block matrix consisting of tridiagonal matrices 
$A_{1},...,A_{p}$ 
and $S$ is the so called spike matrix as given in equation (\ref{eq/spike}), 
where $A_{i}V_{i} = [0\ ...\ 0\ B_{i}]^T$ and $A_{i}W_{i} = [C_{i}\ 0\ ...\ 
0]^T$. 
\begin{equation}
\small DS\hspace{-0pt}=\hspace{-5pt}\begin{bmatrix}
A_{1}\hspace{-5pt}&\\
     &A_{2}\hspace{-5pt}\\      
     &     &\ddots\hspace{-5pt}&\\
     &     &      &A_{p-1}\hspace{-5pt}\\
     &     &      &       & A_{p}
\end{bmatrix}\hspace{-5pt}
\begin{bmatrix}
\hspace{-5pt}I\hspace{-5pt} & V_{1}\hspace{-5pt} \\
W_{2}\hspace{-5pt} & I\hspace{-5pt}      & V_{2}\hspace{-5pt}\\      
      & \ddots\hspace{-5pt} &\ddots\hspace{-5pt}& \ddots\hspace{-5pt}\\
      &        & W_{p-1}\hspace{-5pt} & I\hspace{-5pt} & V_{p-1} \\
      &        &         & W_{p}\hspace{-5pt}  & I\\
\end{bmatrix}
\normalsize
\label{eq/spike}\normalsize
\end{equation}\vspace{-0pt}
The solution to the system then becomes, $DSx = d$ where the system $DY = 
d$ can be used to obtain $Y$, and $Sx = Y$ to obtain $x$. Since matrix $D$ is a 
simple collection of $A_{i}$, each $A_{i}Y_{i} = d_{i}$ can be solved 
independently. Solving $Sx = Y$ requires only solving a reduced penta-diagonal 
system (see Wang et al.~\cite{Wang2014} for a detailed description). The 
algorithm therefore operates in three steps: factorization, reduced system 
solve, and back substitution, where the factorization (LU and UL) has a 
complexity of $O(N)$. The reduced system can be solved directly or indeed can be 
further reduced to a block diagonal system using the truncated-spike variation 
that ignores the outer diagonals when $A$ is diagonally dominant. The Spike 
algorithm is particularly well suited for solving very large systems on 
traditional architectures.
% which we examine and compare to later in Section~\ref{sec/design}.

% Need to use Thomas to factorize the spike matrices and the truncated 
% method to solve the reduced system.

% Contrast to the Spike algorithm (which is also an $O(N)$ algorithm), the 
% Thomas solver 

% Following are the reason not choosing SPIKE :
% 3X Thomas operations on blocks compared to tiled Thomas,
% reduced system is a pentagonal system, hence it requires more computation 
% than 
% tri diagonal reduced system in Tiled Thomas Solver 
% The above is in comparison to our Thomas-Thomas and Thomas-PCR FPGA 
% implementations
% Truncated spike one requires diagonal dominance
% But  our thomas-Thomas or Thomas-PCR can’t solve very large system, due to on 
% chip memory requirement
% FPGA implementation of SPIKE also will hit same issue but truncated spike 
% will 
% work fine for very large systems
\vspace{-0pt}
\subsection{Multiple Tridiagonal Systems in 2D/3D}
% Need to say that our aim is multiple system solves in multiple dimensions -- 
% similar to the argument we did for the TDMA+MPI work
\noindent Each of the above algorithms specifies the solution of a single 
tridiagonal system, which is characteristically a one dimensional problem.  
However, applications of interest are usually 2 or 3 dimensional, where 
tridiagonal systems are formed by solving along one of the coordinate axes. 
This leads to a number of independent systems based on the number of 
discretization points along the other axis. For example a 3D system with 
$N_x\times N_y\times N_z$ number of mesh points will have $N_y\times N_z$ number 
of tridiagonal systems in the first dimension (each system with size $N_x$), $N_x\times N_z$ in the second (each with size $N_y$) and so on. The ADI method, included in the applications we present in this work, repeatedly solve tridiagonal systems along these different axes. Here, the $a_i,b_i,c_i$ and $d_i$ coefficients are calculated for each grid point, in a way that matches the underlying data structure of the application; data is stored contiguously in either a row-major (Z is contiguous, Y, X are strided) or more commonly a column-major (X is contiguous, Y and Z are strided) format. 
This poses a challenge for algorithms that then solve multiple tridiagonal 
systems simultaneously; coefficients for an individual system will be laid out 
differently, depending on the direction of the solve. This is especially true 
on traditional architectures such as CPUs or GPUs~\cite{Laszlo2016}. An FPGA 
design must also carefully consider memory performance when solving such
multi-dimensional applications. 

\vspace{-0.0cm}
\subsection{Related Work}
% Previous TDMA work for FPGAs and their comparison to present work
\noindent Earlier works implementing tridiagonal system solvers on FPGAs such as 
by Oliveira et al.~\cite{ARC2008}, Warne et al.~~\cite{ICCM2012} and Zhang et 
al.~\cite{ReconfigTS2012} used low-level Hardware Description languages (HDL) 
such as VHDL or Verilog for implementing the Thomas algorithm. HDLs require 
extensive hardware knowledge and time/effort in development. These designs were 
restricted to solving 1D or 1D batched tridiagonal systems, instead of full 
multi-dimensional applications. However Oliveira et al.~\cite{ARC2008} 
pipelined both the forward and backward loops and applied data flow between them 
and demonstrated the implementation for a smaller $16^3$ mesh based application 
using only on-chip memory. 
 
With the introduction of High-Level synthesis (HLS) tools, a number of more 
recent works~\cite{ICCS2014, PCTAC2014, ISCAS2015, RC2019} implemented the 
Thomas, PCR, and Spike algorithms on FPGA using HLS tools. Many of these 
works did not demonstrate the solver working on full applications, with the 
exception of L\'{a}szl\'{o} et al. in 2015~\cite{ISCAS2015} which compared a 
one factor Black-Scholes option pricing equation using explicit and implicit 
methods on different architectures such as multi core CPUs, GPUs, and FPGAs. 
The solution by L\'{a}szl\'{o} et al. in 2015~\cite{ISCAS2015} based on the 
Thomas algorithm, targets a Xilinx Virtex 7 FPGA and effectively pipelines both 
forward and backward loops but was not able to apply data flow between these two 
steps. The authors give estimated resource consumption and runtime using Vivado 
HLS for both FP32 and FP64 implementations. Comparing the estimated FPGA runtime 
to a Nvidia K40 GPU shows that the GPU significantly outperforms the FPGA. 

% PCTAC2014, are talking about full spike. but they didn’t implement on 
% FPGA, results are based on HLS estimate and didn’t consider FPGA 
% reconfiguration time

Macintosh, et al. in 2014~\cite{PCTAC2014} uses an OpenCL based implementation 
targeting an Altera Stratix V FPGA using PCR and Spike algorithms. The 
performance on the FPGA is compared to a GPU implementation on an Nvidia Quadro 
4000 GPU. The FPGA performance with PCR is shown to be comparable to that of the 
GPU, but the Spike algorithm on the FPGA outperforms the GPU. Similarly 
Macintosh, et al. in 2019~\cite{RC2019} uses OpenCL to develop 
\texttt{oclspkt}, a library that implements tridiagonal systems solvers 
targeting FPGAs, GPUs, and CPUs. \texttt{oclspkt} uses the truncated spike 
algorithm, and as such will not give exact solutions. However it is able to 
solve tridiagonal systems of any size, taking advantage of interleaved host to 
device transfer to hide the PCIe latency. The work also develops a Thomas 
algorithm based solver that handles larger tridiagonal systems, but does not 
consider pipelining of forward and backward loops. These loops communicate 
through external memory, further limiting achievable performance. Results show 
the FPGA (an Altera Arria 10GX on the Bittware A10PL4 board) performing 
marginally slower than the GPU (a Nvidia M4000) but providing better energy 
efficiency.

The Xilinx quantitative finance library~\cite{xilinxlib} provides a PCR based 
solver, which is a templated implementation for data type, system size steps 
and vectorization. It must be recompiled for different configurations of the 
above parameters. The use of PCR means it requires more FPGA resources due to 
the higher computational intensity of the algorithm. The Xilinx library also 
implements a Douglas ADI solver~\cite{DouglasAndGunn1964} which to our knowledge 
represents a state-of-the-art application implemented with a multi-dimensional 
solver on a Xilinx FPGA.

%Similarly Intel % FPGA library .... need to ref here.

% Need to include a Ref to NAG's intel TDMA work - ask Jacques (this is Intels 
% lib now ?)

% We examine performance of this solver as a baseline in this paper. 

% Xilinx lib seems to be the state-of-the-art so need to say that and provide 
% evidance of how we are better
% Doglus ADI from Xilinx need to be refed here - this is the only multiple 
% system solve and hence the state-of-the-art
% We do batching optimisations and effective pipe-lining or utilising FPGA 
% compute resources effectively when to compare with Xilinx Douglas ADI 

% Compare our implementation at this point
In comparison to above work, the HLS-based synthesis presented in this 
paper, targets the solution of multiple tridiagonal systems and in multiple 
dimensions as commonly found in real-world applications. It uses the Thomas 
algorithm demonstrating that together with techniques such as 
batching~\cite{Kamalakkannan2021} of systems, it provides higher throughput 
for small and medium sized systems. The Thomas algorithm requires a relatively 
smaller amount of DSP resources than the more computationally intensive PCR 
algorithm. For larger systems that do not directly fit in a single FPGA, we 
develop novel Thomas-Thomas and Thomas-PCR solvers to handle a number of 
partitioned systems and then a reduced system solve so that it can operate with 
the available limited on-chip memory of a single FPGA. A further innovation is 
the use of High Bandwidth Memory (HBM) on modern FPGAs which helps to scale the 
design to multiple compute units. To our knowledge, the 2D/3D ADI and SLV 
applications developed in this work, motivated by real-world implicit 
problems on FPGAs is also novel; SLV being one of the few non-trivial 
applications using multi-dimensional tridiagonal solvers presented in 
literature. The Thomas based solver developed in this paper gives higher 
performance than the current PCR based Xilinx library, as we will show in 
Section~\ref{sec/perf}. The Thomas algorithm is better suited 
for obtaining high-throughput when solving batches of tridiagonal systems than 
PCR. Douglas ADI solver from Xilinx is also based on PCR. Thus, it would be 
similarly less performant, although implementing a different numerical method, 
than the ADI solvers in our work. Additionally, the predictive analytic model 
and the performance comparison with a state-of-the-art GPU based tridiagonal 
solver library gives a much needed frame of reference for evaluating our FPGA 
design's performance, providing insights into the feasibility and profitability 
of an FPGA design for realistic workloads.

\vspace{-0.2cm}
\section{FPGA Design}\label{sec/design}
\vspace{-0.1cm}

\noindent An FPGA use a multiple-instructions, single data (MISD) architecture 
to implement computation, be it a single kernel or a series of kernels. There 
is no fixed general purpose architecture that can be exploited using software as 
a traditional CPU or GPU does. Instead a fixed circuit of the computation is 
synthesized using a variety of basic circuit elements. These are digital signal 
processing (DSP) blocks to implement arithmetic operators, look-up-tables (LUTs) 
and registers, fast on-chip block memories (BRAM/URAM), clock modules, and a 
rich routing fabric to connect these elements into a large logical architecture. 
The overall die consists of a number of these which are called Super Logic 
Regions (SLR). The U280 has 3 SLRs. Bandwidth within an SLR is extremely high 
(TB/s) due to the wealth of connections and memory elements, while between SLRs 
it is limited by the number of silicon connections available. An FPGA board will 
also include much larger, but slower DDR4 (32 GB on the U280) memory as external 
memory. Managing the movement of data between these different types of memory is 
key to achieving high computational performance. The introduction of High-Level 
Synthesis (HLS) tools has reduced the complexity of FPGA programming, where a 
high-level programming language such as C++/OpenCL can be used with special 
directives to target the FPGA. However, getting good performance is still 
significantly challenging as code needs to be structured to suite the 
data-flow/pipelined programming style. The key optimizations required to obtain 
the best performance are transformations enabling pipelining, replication of 
circuit units (CUs) and tiling to improve locality such that data can be reused 
by fitting to fast on-chip memory. For a good overview of these techniques we 
refer the reader to the paper by De Matteis et al.~\cite{DeMatteis2020} and the 
Xilinx HLS programming 
guide~\cite{Xilinxswmanual2020}.

% Actual design from this work
Considering the resources available on an FPGA, a single tridiagonal 
system solve, using the Thomas algorithm in Algorithm~\ref{alg/thomas}, would 
require 4 multiplications, 1 division and 2 subtractions for the forward path 
(lines 3-7) and one multiplication and subtraction for the backward path (lines 
8-10). However, given that there are dependencies for computing $d_{i}^{*}$ and 
$c_{i}^{*}$, each iteration of the forward path loop will have to be executed 
serially, incurring the full arithmetic pipeline latency, $l_{f}$ ($\approx$30 
clock cycles on a Xilinx U280 FPGA), to pass through the forward loop datapath. 
Additionally the backward loop can only start when all iterations of the 
forward path have been completed, due to the reverse data access where the loop 
starts from iteration $N-2$. Thus the total latency for solving a single system 
with the Thomas algorithm would be approximately $l_{f}\times N + l_{b}\times 
N$ clock cycles (assuming $l_{b}$ cycles is the arithmetic pipeline latency for 
completing a single iteration of the backward loop). On the other hand, a PCR 
based single solver implementation would require 4 subtractions, 9 
multiplications and 1 division within the inner loop of Algorithm~\ref{alg/pcr}. If 
$l$ is the arithmetic pipeline latency of the inner loop, then the total 
number of clock cycles for the PCR algorithm, is $(N+l)\times logN$. Here we 
assume that the outer loop is executed serially. Given the inner loop 
iterations are independent, they can be unrolled by some factor $f_{U} = 
2,3,...$ which will then require $f_{U}\times$ the resources to implement the 
inner loop. The total clock cycles will then be $(N/f_{U}+l) \times logN$. The 
outer loop iterations have a dependency and thus cannot be unrolled. 

For the Thomas solver, there are $l_f$ clock cycles between consecutive 
iterations of a single system solve in the forward path. This can be considered 
as a \textit{dependency distance}. As such, we could attempt to solve $l_f$ 
number of tridiagonal systems to fully utilize the forward path circuit 
pipeline. This can be done by \textit{interleaving} the iterations of the 
forward pass loop of of the Thomas solver such that iteration 1 of system 1 is 
input followed by iteration 1 of system 2 and so on, per clock cycle, up to 
iteration 1 of system $l_f$. In fact selecting a group, $g = MAX(l_f, l_b)$ 
enables $g$ system solves to be interleaved, saturating the pipeline. If there 
are $B$ number of total tridiagonal systems to be solved, i.e. a batch size of 
$B$, then the total latency with Thomas can be shown to incur a latency given 
by (\ref{eq/batchedthomas}):\vspace{-0.0cm}
\begin{equation}(1 + \ceil*{B/g})\times gN 
\label{eq/batchedthomas}\vspace{-0.1cm}\end{equation}
Thus for large $B$ the total latency tends to be $BN$. This is a characteristic 
of all $O(N)$ algorithms, which ideally can be pipelined to take input each 
clock cycle at the cost of differing resource consumption.

For the PCR algorithm, given there are no dependencies between iterations of a 
single system, solving a batch of $B$ systems (by batching the inner loop) 
incurs the latency in~(\ref{eq/batchedpcr}): 
\begin{equation}(BN/f_{U}+l)\times logN \label{eq/batchedpcr}\end{equation}
For large $B$, dividing (\ref{eq/batchedpcr}) by (\ref{eq/batchedthomas}) gives 
a factor of $logN/f_{U}$ pointing to the fact that the batched Thomas solver is 
$logN$ times faster than batched PCR, for $f_{U} = 1$. Thus, to match the 
Thomas solver latency, a batched PCR implementation needs an unroll factor 
$f_{U} = logN$.  However, given that the PCR inner loop has a considerably 
larger resource requirement, compared to the Thomas solver, on a given FPGA with 
fixed amount of resources, the batched Thomas solver will always provide better 
performance. The exception occurs when the system size, $N$ is large 
and FPGA on-chip memory becomes the limiting factor. We discuss 
the design and best algorithms for such cases in Sec~\ref{subsec/largesystems}.

% Thomas = l_f*N + ceil(B/l_b) * l_b * N

% Need a diagram illustrating the data access patterns in z-dims 

\begin{figure*}[h]
\begin{subfigure}{.5\textwidth}
\centering
\includegraphics[width=7cm]{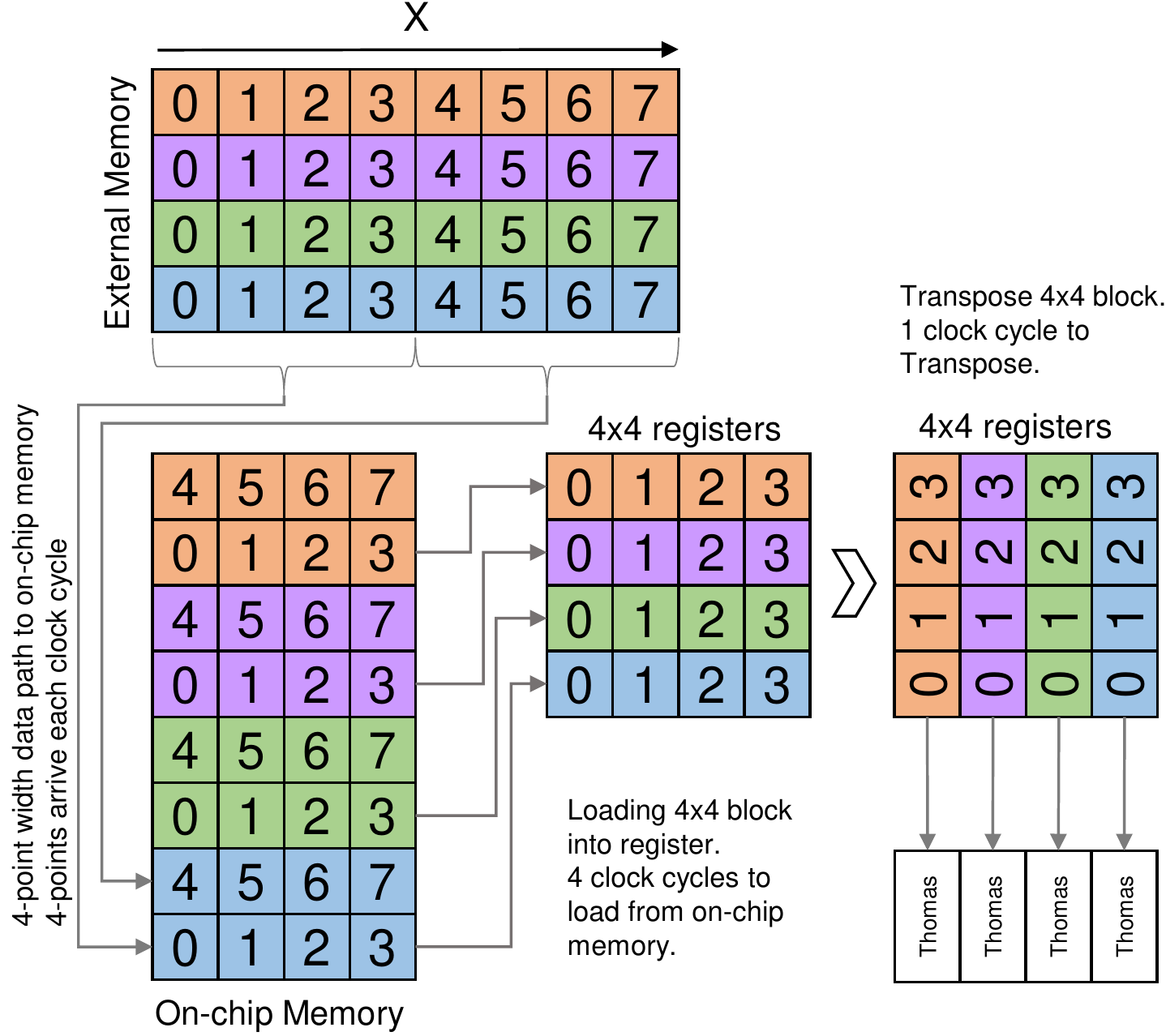} \vspace{-0.1cm}
\caption{x-dim solves}
\label{fig/transpose}
\end{subfigure}\vspace{.4cm}
\begin{subfigure}{.5\textwidth}
\centering\vspace{.3cm}
\includegraphics[height=5.5cm]{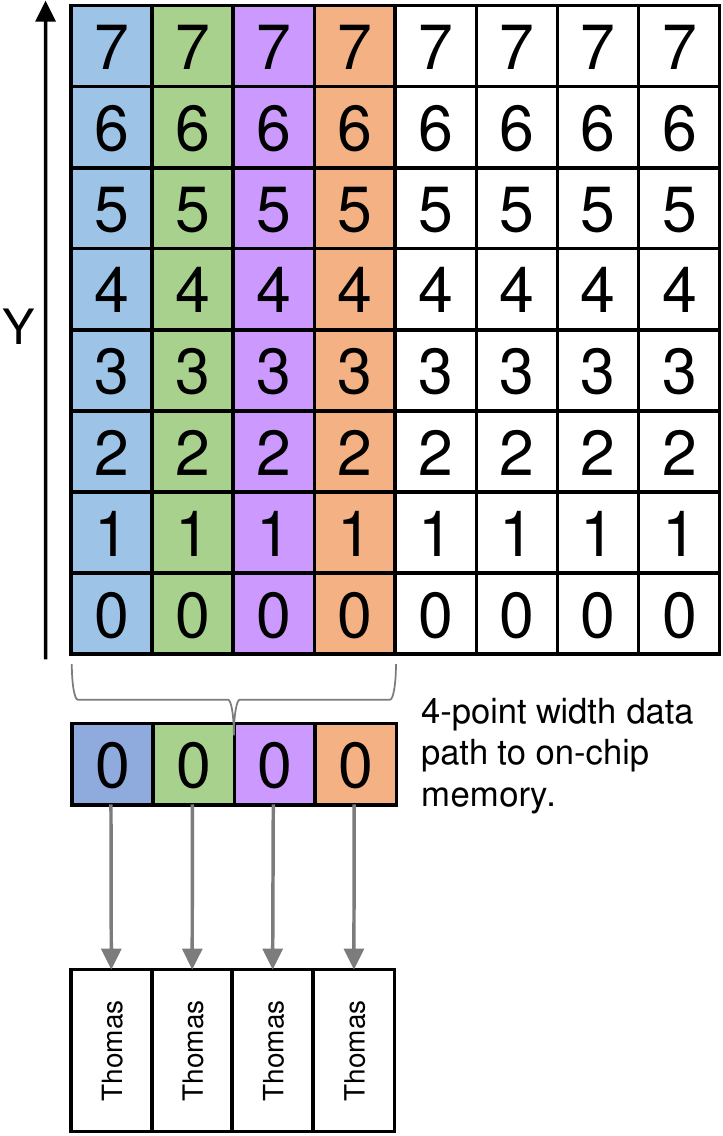} \vspace{.2cm}
\caption{y-dim solves}
\label{fig/ydimsolve}
\end{subfigure}\vspace{-0.5cm}
\caption{Datapath for x- and y-dim solves. Illustrating a 4-point data 
path width and $4\times$ (vectorized) Thomas solvers.}\vspace{-0.6cm}
\label{fig/datapath}
\end{figure*}

Considering a batched solver based on the spike algorithm, assume each 
system in the batch is of size $N$.  The algorithm creates $m$ number of 
blocks and each has LU and UL factorization done in parallel, followed by the 
penta-diagonal solve and then back-substitution in parallel. This incurs
a total latency given by (\ref{eq/batchedspike}) :
\begin{equation} (1 + \ceil*{Bm/g+1})\times gN/m + mC  
\label{eq/batchedspike}\end{equation}
The latency for the factorization for each block (first term), is similar to a 
Thomas forward and backward solve carried out in an interleaved manner. 
Although the number of cycles spent on the pentadiagonal reduced system 
solve is $BmC$ (assuming a linear latency model) and back substitution is $BN$, 
only the latency for first stage of pentagonal solver is added to equation 
\ref{eq/batchedspike} as all three modules are pipelined. Back substitution does 
not add additional delay between its inputs and outputs. When $B$ is 
sufficiently large and stages are pipelined, a latency of $BN$ is achieved.
% , assuming $C/m < 1$ for the pentadiagonal solver. 
Again this is due to the spike algorithm having a $O(N)$ complexity. However, if 
$BmC \geqslant BN$ then data flow must stall for some time decreasing 
throughput. Considering resource consumption the LU/UL factorization requires 
$3\times$ the resources for an equivalent Thomas solver and the pentadiagonal 
solver requires additional resources (again more than an equivalent Thomas 
solver). \\
\indent Given the lower resource requirements and profitability of the Thomas 
algorithm, compared to the other algorithms, we first focus on its optimized 
batched implementation on an FPGA for system sizes that can fit 
into on-chip memory. As we are interleaving groups of $g$, the $a_{i-1}, 
b_{i-1}$ and $c_{i-1}$ values needs to be stored in on-chip memory such that 
they can be used in subsequent ($i$th) iterations. For a FP32 implementation we have 
found that a grouping of 32 is sufficient to effectively pipeline the 
computation (this is 64 for FP64) on the Xilinx Alveo U280. The forward and backward 
loops operate in opposite directions and thus a First-In-First-Out (FIFO) buffer 
cannot be used, rather on-chip addressable memory is used for data movement. The forward 
and backward loops can be made to operate in parallel when batching a number of 
system solves, using ping-pong buffers (also called double buffers). With this 
technique, dual port memory is partitioned into two parts, one for the writing 
process the other for the reading. Once writes (by the forward pass) and reads 
(backward pass) are completed, read and write locations are swapped. Note that 
the very first read has to wait till the very first write has completed. 
Additionally, the technique also doubles the memory requirement compared to 
using the same memory portion for both read and write. Including the latency for 
starting the first write to the ping-pong buffer, and writing back the final 
result to external memory increases the total latency in 
(\ref{eq/batchedthomas}) by $2gN$ to give: $(3 + \ceil*{B/g})\times gN$ clock 
cycles. The total on-chip memory required for a single Thomas solver 
interleaving $g$ systems can be computed based on the need to store the 
$a,c,d,c^{*},d^{*}$ and $u$ vectors, where each consumes $2gN$ words in the 
ping-pong buffers. The total $12gN$ requirement with dual port memory can be 
satisfied with 6$\times$ dual port block RAMs (URAM/BRAM) each with a capacity 
of $2gN$. Additionally there is a need to store $g$ values of the $i-1$th 
iteration separately, requiring 4 RAMs with a capacity of $g$ words.\\
\indent Data transfer from external memory to on-chip memory plays a crucial 
part in achieving high performance, especially for multi-dimensional solvers 
such as the 3D ADI heat diffusion application detailed later in this paper. If 
we consider a 3D application with systems sizes ($N$) of 256 in all three 
dimensions, then a solve along the x-dimension will have YZ ($256\times256$ in 
this case) number of systems to be solved, each corresponding to an 
\textit{x-line} system of size $256$. Given the data is stored in consecutive 
memory locations along the x-lines, good memory throughput can be achieved. 
However to exploit the full memory bandwidth, a larger number of memory ports needs 
to be used. For the 512-bit memory ports, on the Alveo U280, it is sufficient to 
saturate the data-flow pipeline with a width of 256-bits at a 300MHz clock speed, 
which is our target frequency. This enables us to fetch data sufficient to feed 
8 Thomas solvers in parallel. Such a configuration can be viewed as a 
\textit{vectorized} Thomas solver. Additionally, the total YZ number of x-lines 
can be set up to be solved in groups ($g$) of 32. Here, the 1st Thomas solver 
circuitry solves the 0th, 8th, 16th and so on x-lines, the 2nd solves 1st, 9th, 
17th and so on x-lines, and so on. Batches of x-lines can be solved in such 
interleaved groups to saturate the data flow pipeline to achieve higher 
throughput. 

In the x-dimension, the reads from external memory bring in data stored in 
consecutive memory locations. However, the data fetched belongs to the same 
line (i.e. same system), thus we need to buffer 8 x-lines internally and 
carry out an $8\times8$ transpose to feed that to 8 different solvers (see 
\figurename{~\ref{fig/transpose}} for an illustration of the issue with a
$4\times4$ transpose). For solving along the y-dimension, we fetch 
each XY plane to on-chip memory to avoid strided memory accesses and then 
read along the y-lines from the on-chip memory (see 
\figurename{~\ref{fig/ydimsolve}}). Similarly for solving along the 
z-dimension, we read in x-lines (which are consecutive in memory) along the z 
dimension, fetching XZ planes, to on-chip memory. No transpose is required for 
y- and z-dimension solves as each element corresponds to a different system. 
Utilizing the HBM available on modern FPGAs, the full vectorized Thomas solver, 
which can be viewed as a single compute unit (CU), can be instantiated a number 
of times to obtain further parallel performance. Specific designs for 
applications with multiple CUs are discussed in Section~\ref{sec/perf}. For a 3D 
application, the x- and y-dimension solves can be effectively pipelined, storing 
the resulting XY planes in on-chip memory without writing to external memory. 
However the z-dimension solve requires reading from external memory. As such 
2D applications can be further optimized with unrolling. Again we will discuss 
specific implementations with unrolling in Section~\ref{sec/perf}.

% (7) Finally also need to compare with xilinx lib (one liner ?)

% (B/m+3)*m*N  - but in the full Thomas design, we assume data is from the 
% FIFO,  then we have to put into the memory and at final we have to put back to 
% FIFO  from memory hence there will be additional 2*mN latency

% DONE(1) talk about B,C,D storing for next iter on on-chip memory - HLS 
% dependency issue, grouping (32 systems for FP32) for interleaving and then 
% batching 
% DONE(2) Data flow between FW and BW loop - should use onchip memory for data 
% movement, ping-pong buffer
% DONE(3) Full Thomas flow

% DONE(4) Vecotorization, Multiple CUs and HBM memory use
% DONE(6) Handling multi-dim tridiagonal solver problems - what needs to be 
% done for performance
% DONE (5) Resource model for full Thomas for overall implementation - see 
% slide 63
% (7) Finally also need to compare with xilinx lib (one liner ?)

% Thomas will always best when there is multiple systems and we are not limited 
% by on chip memory. But we will start to hit on chip memory limitation when 
% system size goes around 128-256. Then Thomas-PCR becomes better than Just 
% Thomas

% \noindent 
% 1D Thomas (FP32)\\ 
% 32 Systems with Thomas (FP32)\\
% 32 Systems with Thomas with batching (FP32)\\
% Vectorization (idea) - Optimization to obtain more parallel execution\\
% multiple CUs description  + HMB memory\\

\vspace{-5pt}
\subsection{Larger Systems Solve}\label{subsec/largesystems}
\begin{figure}[t]
\centering
\includegraphics[width=7cm]{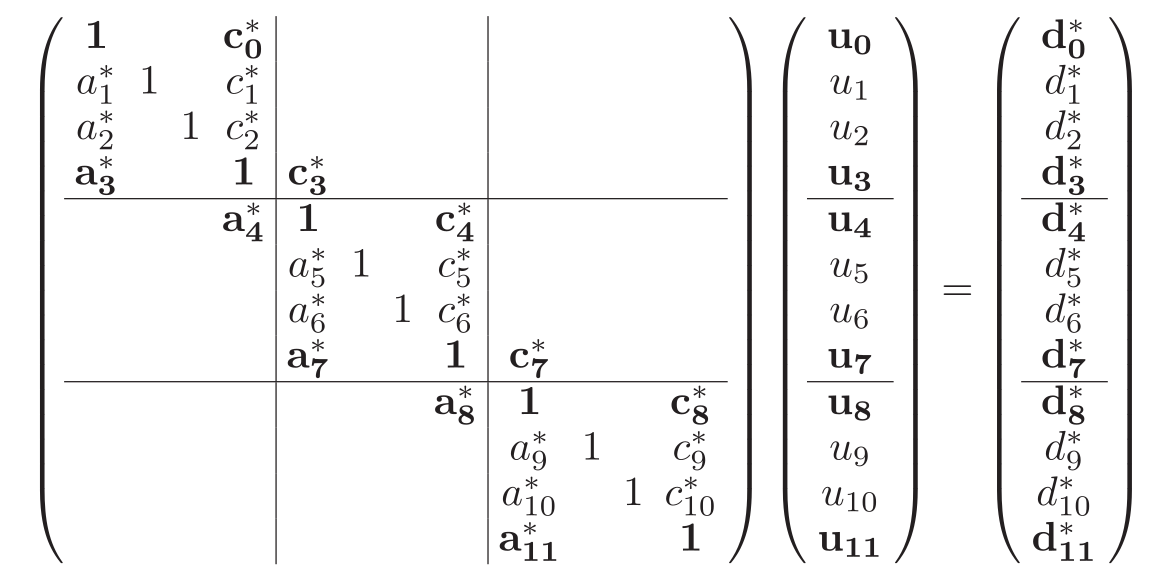} \vspace{-0.3cm}
\caption{Splitting to three subsystems - unknowns after the forward and 
backward pass of a modified Thomas solver}\vspace{-0.6cm}
\label{fig/modifiedthomas}
\end{figure}

\noindent Interleaved solving of systems require on-chip memory proportional to 
the system size, $N$ and number of groups $g$. As such the maximum size of the 
system that can be solved is limited by the FPGA on-chip memory resources. 
We can split the tridiagonal system into subsystems (or \textit{tiles}) of size 
$M$ where each subsystem can be solved using a modified Thomas solver, where 
after a forward and backward phase, each unknown is expressed in terms of 
two unknowns $u_o$ and $u_{M-1}$: \vspace{-0.3cm}
\begin{equation}
a_{i}u_{0}+u_{i}+c_{i}u_{M-1} = d, \quad i = 1, 2, ..., M - 2 
\end{equation}
This results in a reduced tridiagonal system spread across each sub-domain as can 
be seen in Fig~\ref{fig/modifiedthomas} (as detained by L\'{a}szl\'{o} 
et al.~\cite{Laszlo2016}). The unknowns at the beginning and end of each 
subsystem can be solved again using the Thomas algorithm, or indeed PCR. 
Finally, the result from the reduced system, is substituted back into the 
individual subsystems (see L\'{a}szl\'{o} et al.~\cite{Laszlo2016} which 
implements a Thomas-PCR solver for GPUs).

The \textit{tiled-}Thomas-Thomas solver requires additional circuitry implemented 
to solve the reduced system, but larger systems can be solved. To achieve 
higher performance, forward and backward phases over tiles can be interleaved. 
The reduced system size $N_{r}$ is double the number of tiles. Solving the 
reduced system with Thomas requires $2gN_{r}$ clock cycles. This should not 
exceed the clock cycles taken by the forward and backward phases over the tiles. 
At the end of the backward phase, results ($a^*, c^*$ and $d^*$) are stored in 
a FIFO buffer while the reduced system for each tile is computed. Then the 
reduced system results can be substituted back to complete the solve. Using a 
FIFO maintains the data-flow pipeline without stalling. 

Considering a system of size $N$, split into $t$ number of tiles (note then 
$N_r = 2t$), assume we interleave $g$ number of tiles using the Thomas-Thomas 
algorithm to solve a total number of $B$ systems. Then the total latency is 
given by (\ref{eq/thomasthomas}): \vspace{-0.1cm}
\begin{equation} (2 + \ceil*{Bt/g})\times \ceil*{N/t} g + 
g_{r}\times(2t)\times2 \label{eq/thomasthomas}\vspace{-0.0cm}\end{equation} 
The second term is for the reduced solve. The $g_r$ is similar to $g$, but it is 
equal to or larger than number of interleaved systems for the reduced solve. It 
is 32 for FP32 and 64 for FP64 on the U280. Similarly, based on the latency for 
solving the first phase of the algorithm on a tile, the number of systems to be 
interleaved is $\ceil{32/t}$ for FP32 and $\ceil{64/t}$ for FP64. For larger B, 
we can see that the latency tends to $Bt\ceil*{N/t}$. Considering on-chip memory 
requirements the forward and backward phases of the modified Thomas can be shown 
to require $9\times2\times g/t \times N$ words that can be satisfied by 9 RAMs 
setup as ping-pong buffers. Here we note that larger $t$ lead to lower memory 
requirement. The reduced solve requires much less memory, 
$7\times2\times2t\times\ceil{g/t}$ in the form of 7 ping pong buffers. 
Furthermore, a FIFO buffer space would be required, equivalent to the maximum 
clock cycles spent on the reduced system, as we have to flush each point per 
clock cycle from the backward phase. 

% ($2\times g_{r}/t \times 2t$ + $g_{r}/t$ words in total). 

The reduced system solve can also be implemented with the PCR algorithm 
resulting in the latency given in (\ref{eq/thomaspcr}). 
\vspace{-0.0cm}
\begin{equation} (2 + \ceil*{Bt/g})\times \ceil*{N/t} g + (2t+l)\times log(2t) 
\label{eq/thomaspcr} \vspace{-0.0cm}\end{equation}
Again for larger B, this tends to $Bt\ceil*{N/t}$, however, there is a lower 
on-chip memory requirement of $(2t+l)\times log(2t)$ words for each of $3$ FIFO 
buffer, due to the lower latency for reduced system solve in PCR. Since data 
flow design requires matching performance of solving tiles and the reduced 
system and as PCR is faster when solving reduced systems, the number of tiles 
can be increased even for smaller systems, further reducing requirement for on 
chip memory for the first phase of the algorithm. As such we can conclude that 
the Thomas-PCR version would result in better performance. 

% 32 or 64 above is the core circuit latancy for phase 1 of modified Thomas

% DONE Need performance model in clock cycles for Thomas-Thomas
% DONE Need resource model for Thomas-Thomas
% DONE Latancy for Thomas-PCR
% DONE Resource Model for Thomas-PCR
% Details of selecting Thomas or Thomas-Thomas based on system size - tabular 
% method

% This is fairly novel only the Australian paper is about larger systems, but i 
% am not sure they have done all the optimisations. they didn’t provide details 
% of the implementation -- we might want to add this to the novel contributions

% Hybrid approach is suitable for larger meshes and we know roughly what size 
% it is going to be profitable, below this size the pure Thomas with batching 
% will be superior. The stalling argument will also support this.

% \subsection{Performance Model}\label{subsec/model} 
% This need not be a subsection
% Summary table of the performance model. Both small/mediam and large mesh 
% tridiagonal solver implementations including resource models

% We first compare performance of our library to the existing Xilinx library 
% based on the PCR algorithm. 

\begin{figure}\centering
\includegraphics[width=6.5cm]{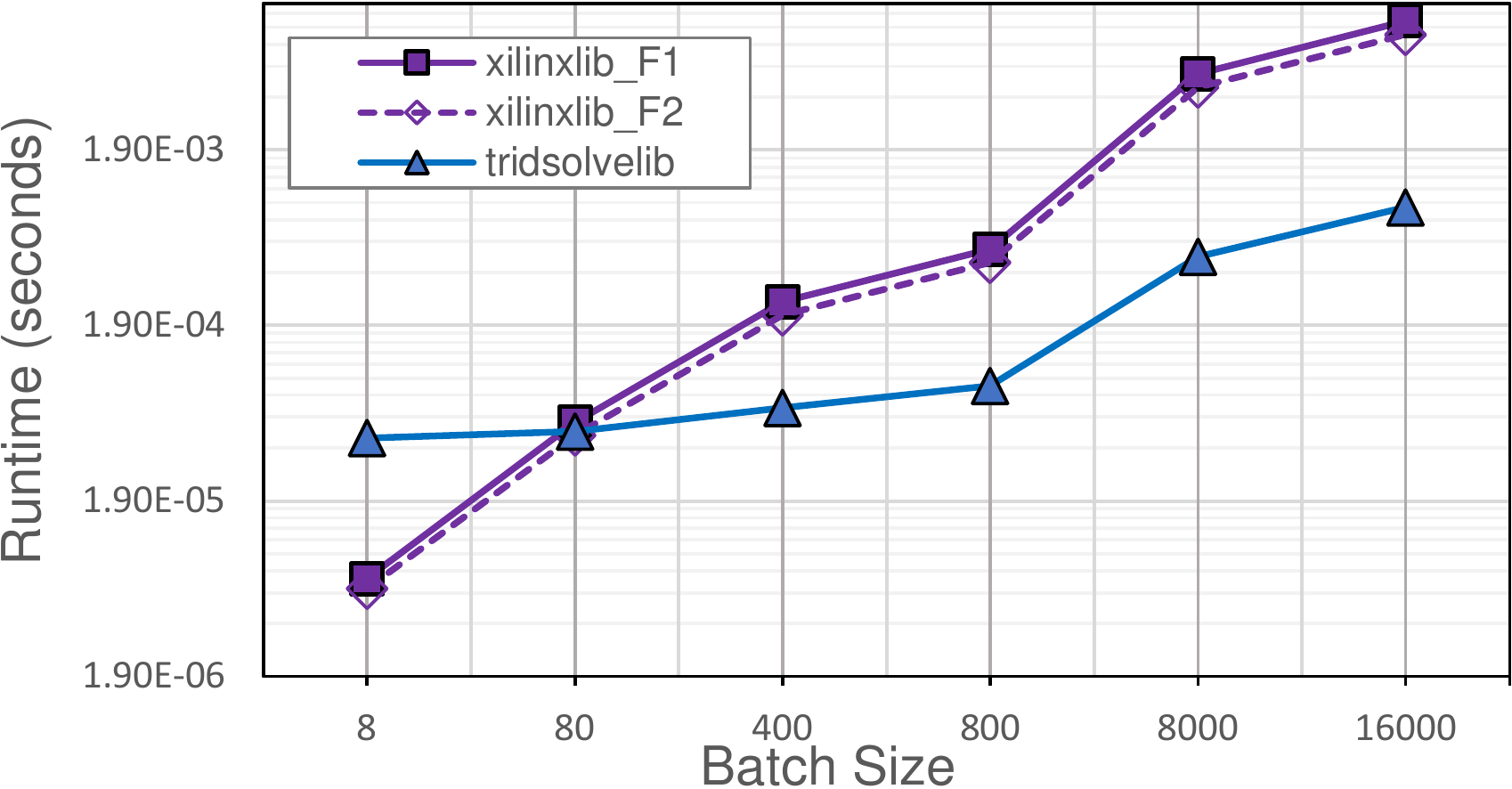}
\vspace{-0.2cm}\caption{\small \texttt{tridsolvlib} vs \texttt{xilinxlib}, System 
size-128}\vspace{-0.0cm}\label{grp/xilinxlib}\vspace{-0.55cm}
\end{figure}

\vspace{-0.3cm}
\section{Performance}\label{sec/perf}
\vspace{-0.1cm}

\noindent In this section we present performance from our FPGA design strategy 
for high-throughput tridiagonal system solvers. First, we briefly compare the 
performance of our library to a current state-of-the-art FPGA tridiagonal solver 
library from Xilinx~\cite{xilinxlib} which is based on PCR, demonstrating the 
higher performance gains from a batched Thomas-based solver as predicted by the 
analytic model developed in Section~\ref{sec/design}. Batching of systems is key 
to higher performance. \figurename{~\ref{grp/xilinxlib}} presents performance 
of 1D tridiagonal systems of size 128, solved using the Xilinx library 
(\texttt{xilinxlib-F1}) compared to our Thomas algorithm-based implementation 
(\texttt{tridsolvlib}) on a range of batch sizes. As discussed in the 
analytic model, for larger batch sizes the Xilinx library performed 
significantly slower than the Thomas based solver. Adding further optimizations, 
such as inner loop unrolling
%, batching of inner loop 
and a FIFO data path to the 
Xilinx solver (\texttt{xilinxlib-F2}) only marginally improved performance, 
leaving an order of a magnitude performance gap. We also observed that the 
PCR-based \texttt{xilinxlib-F2} implementation consumed higher resources (an 
example is given in Table~\ref{tab/xilinxlib} for the batch size of 8000 
systems).

% Table~\ref{tab/xilinxlib} details results for solving a batch of 8000, 1D 
% systems of size 128 using the Xilinx library (\texttt{xilinxlib-F1}) compared to 
% our Thomas algorithm-based implementation (\texttt{tridsolvlib}). The Xilinx 
% library performed significantly slower than the Thomas based solver. Adding 
% further optimizations, such as inner loop unrolling, batching of inner loop and 
% a FIFO data path to the Xilinx solver (\texttt{xilinxlib-F2}) only marginally 
% improved performance, leaving an order of a magnitude performance gap. We also 
% observed that the PCR-based \texttt{xilinxlib-F2} implementation consumed higher 
% resources. 

% \begin{comment}% Version 1
In the remainder of this section we focus on using our FPGA design strategy. 
Specifically, we apply it to two representative, non-trivial applications. We 
investigate both 2D and 3D applications, with both FP32 and FP64 
representations. Model predicted resource utilization estimates are used to 
determine initial design parameters and model predicted runtimes, which we 
compare to actual runtimes of the applications on a Xilinx Alveo U280. We use 
Vivado C++ due to ease of use for configurations and support of some C++ 
constructs compared to OpenCL. However, we note that OpenCL could equally be 
used to implement the same design. Finally, we compare performance on the FPGA 
to an NVIDIA Tesla V100 GPU using the tridiagonal solver library, 
\texttt{tridsolver} implemented by L\'{a}szl\'{o} et 
al.~\cite{Laszlo2016}~\cite{tridslv} using its batched version presented by 
Reguly et al.~\cite{RegulyBatching2019}. This GPU library has been 
shown~\cite{Balogh2021} to provide matching or better performance than the two 
current batch tridiagonal solver functions -- 
\texttt{cusparse<t>gtsv2StridedBatch()} and 
\texttt{cusparse<t>gtsvInterleacedBatch()}, in Nvidia's cuSPARSE 
library~\cite{cusparse,Valero-Lara2017}. Our experiemnts also confirmed these 
results for the applications evaluated in this paper. Additionally it 
features direct support for creating multi-dimensional solvers, whereas 
cuSPARSE requires data layout transformations, for example in between doing an 
x-solve and a y-solve to implement multi-dimensional problems. Thus we use 
\texttt{tridsolver} in our evaluation throughout this paper, but note that 
cuSPARSE libs would have equally provided the same insights when compared to 
the FPGA solvers on the U280. Given that previous work has demonstrated GPUs to 
provide significantly better performance than multi-threaded 
CPUs~\cite{Laszlo2016}, we do not compare with CPU runs. Table~\ref{tab/systems} 
briefly details the specifications of the FPGA and the GPU systems (both 
hardware and software) used in our evaluation. The V100 is based on 12nm gate 
size comparable to the U280's 16nm gate size. It has a peak bandwidth of 
900GB/s, nearly twice that of the U280's 460GB/s bandwidth. 

\begin{table}[t]\footnotesize
\centering
\caption{\small Xilinx library performance : 8000 systems of size 
128\normalsize}\vspace{-5pt}
\setlength\tabcolsep{0.1cm}
\begin{tabular}{@{}llllll@{}}
\toprule
Design                  &Runtime (ms) & BW (GB/s) & DSP & URAM & BRAM \\
\toprule
\texttt{tridsolvlib}    & 0.47        & 43.34     & 218 & 76   & 50   \\\hline
\texttt{xilinxlib-F1}      & 5.15        &  3.97     & 447 & 20   & 70  \\\hline
\texttt{xilinxlib-F2}      & 4.32        &  4.73     & 655 & 20   & 102  \\
% \texttt{xilinxlib-opt}  & 4.17        &  4.90     & 251 & 20   & 70   \\
\bottomrule
\end{tabular}\label{tab/xilinxlib}\vspace{-0.2cm}
\end{table}\normalsize

\begin{table}[t]\footnotesize
\centering
\caption{\small Experimental systems specifications.\normalsize}\vspace{-0.2cm}
\begin{tabular}{@{}ll@{}}
\toprule
FPGA	      & Xilinx Alveo U280~\cite{u280}                \\
\midrule
DSP blocks    & 8490 \\
BRAM / URAM   & 6.6MB (1487 blocks) / 34.5MB (960 blocks)\\
HBM           & 8GB, 460GB/s, 32 channels\\
DDR4          & 32GB, 38.4GB/s, in 2 banks\\
Host	      & AMD Ryzen Threadripper PRO 3975WX (32 cores)\\
              & 512GB RAM, Ubuntu 18.04.6 LTS    \\
Design SW     & Vivado HLS, Vitis 2019.2 \\
\toprule
GPU	      & Nvidia Tesla V100 PCIe~\cite{v100}      \\
\midrule
Global Mem.   & 16GB HBM2, 900GB/s \\
Host	      & Intel Xeon Gold 6252 @2.10GHz (48 cores)\\
              & 256GB RAM, Ubuntu 18.04.3 LTS    \\
Compilers, OS  & nvcc CUDA 10.0.130, Debian 9.11 \\
\bottomrule
\end{tabular}\label{tab/systems}\vspace{-0.4cm}
\end{table}\normalsize

\begin{figure*}[t]\centering
\centering
\subfloat[][\centering\footnotesize 2D FP32 - 120 iter.$v=8,f_U=3,N_{CU}=3$]
{\vspace{-0.2cm}\includegraphics[width=6.5cm]{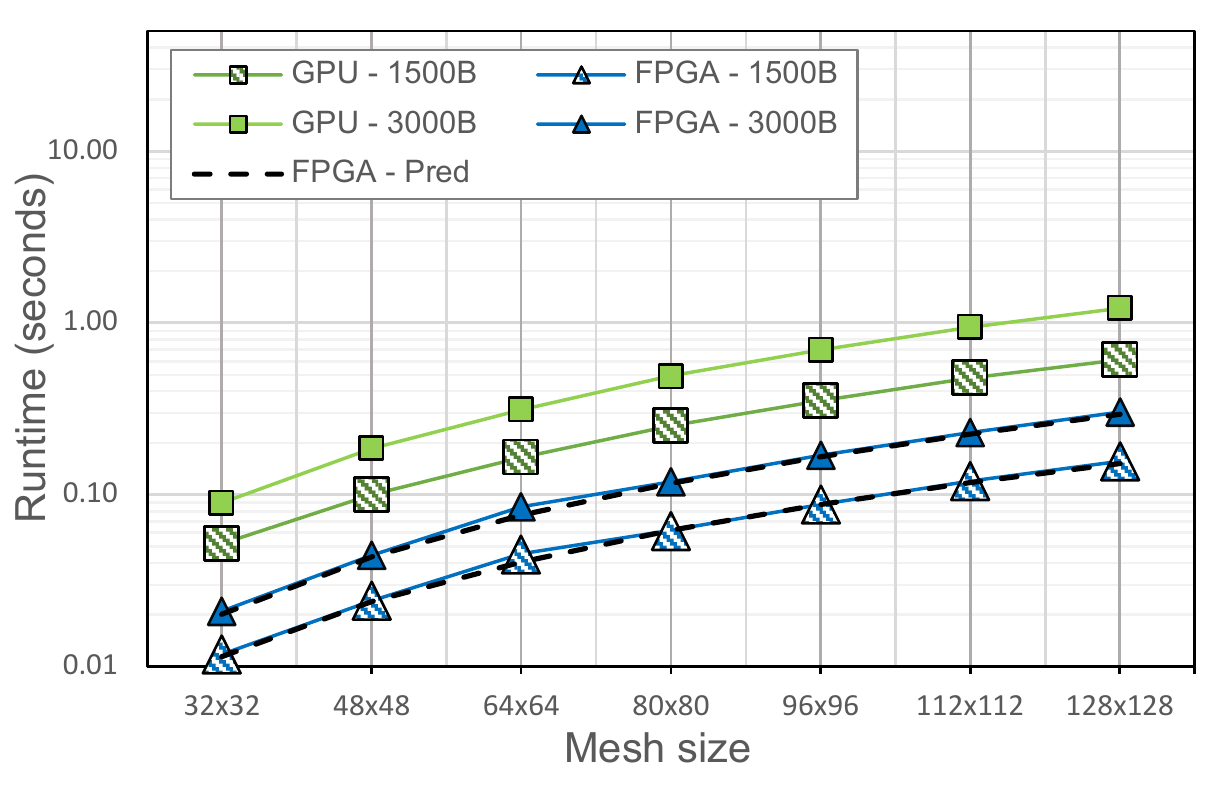}}\hspace{
0.4cm}
\subfloat[][\centering\footnotesize 2D FP64 - 120 iter.$v=8,f_U=2,N_{CU}=3$]
{\vspace{-0.2cm}\includegraphics[width=6.5cm]{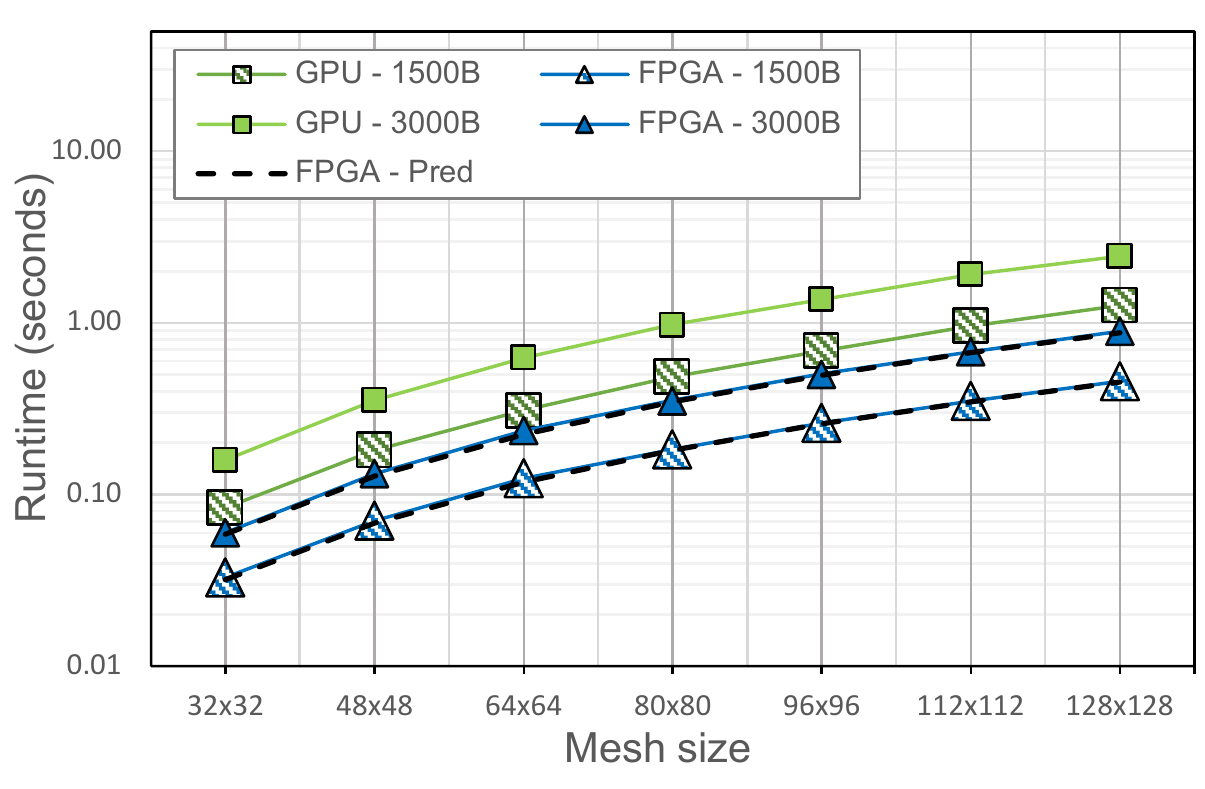}}\hspace{
0.4cm}
\vskip\baselineskip
\subfloat[][\centering\footnotesize 3D FP32 - 100 iter.$v=8,N_{CU}=6$]
{\vspace{-0.2cm}\includegraphics[width=6.5cm]{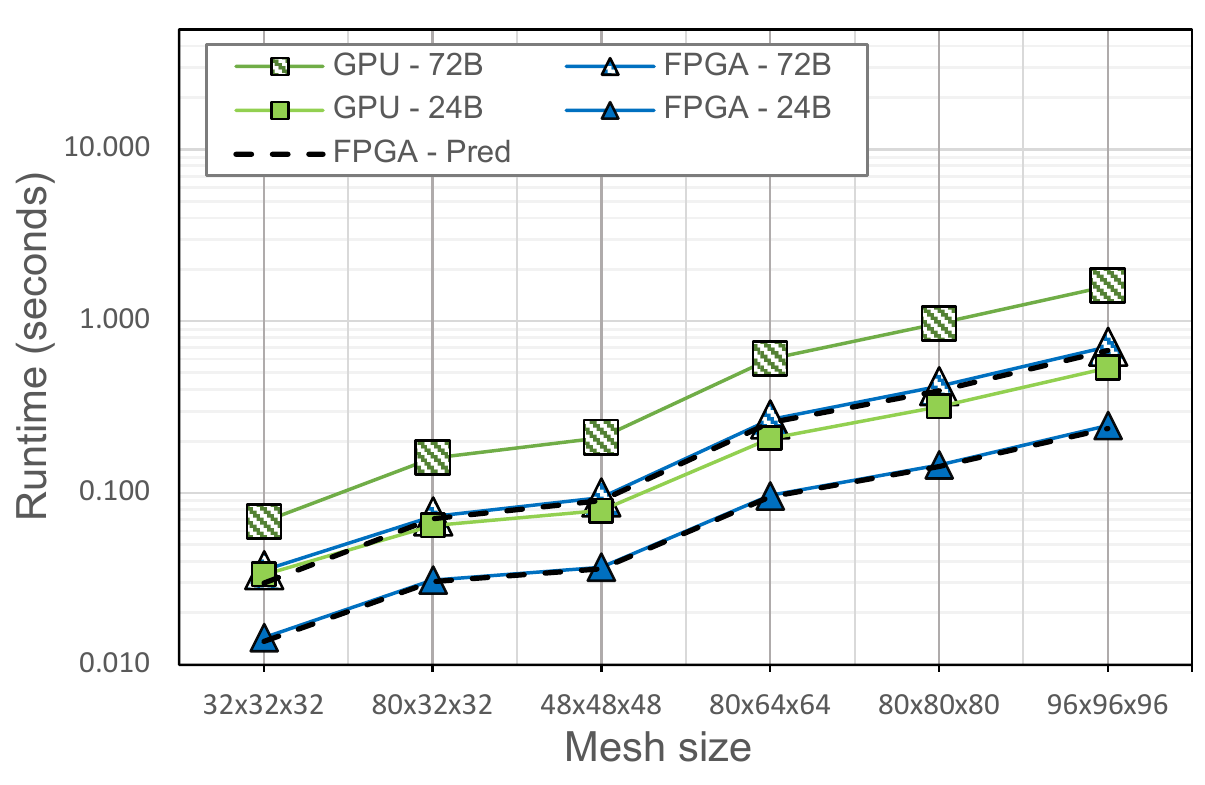}}\vspace{
-0.2cm}\hspace{0.4cm}
\subfloat[][\centering\footnotesize 3D FP64 - 100 iter. $v=8, N_{CU} = 3$] 
{\vspace{-0.2cm}
\includegraphics[width=6.5cm]{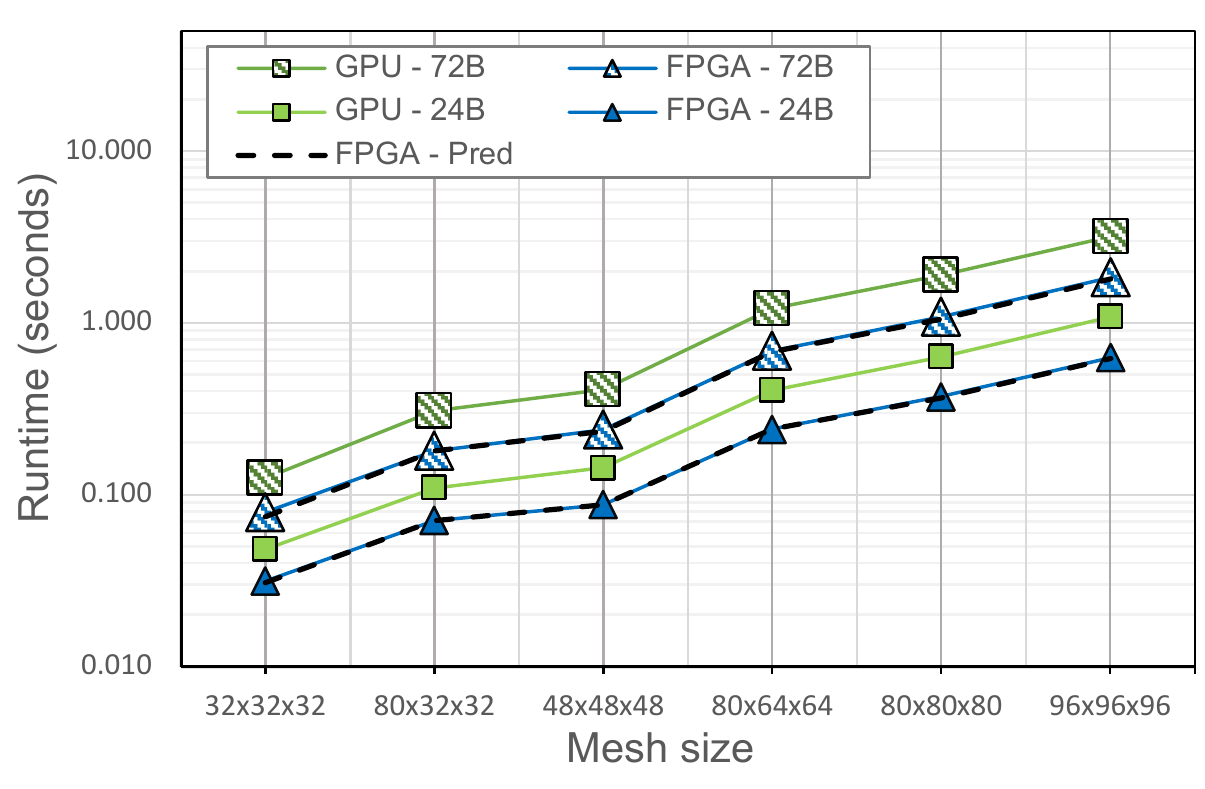}}\vspace{-0.0cm}\hspace{0.4cm
}
\caption{\small ADI Heat Diffusion Application 
Performance}\vspace{-0.6cm}\label{grp/ADI}
\end{figure*}

\vspace{-0.2cm}
\subsection{ADI Heat Diffusion Application}\vspace{-0.1cm}
\noindent The first application is an ADI based solve of the heat diffusion
equation. The high-level algorithm of the application in 3D is detailed in 
Algo.~\ref{alg/adi}. The application consists of an iterative loop 
which starts by calculating the RHS values using a 7-point stencil, followed 
by calls to the tridiagonal solver for each of two or three dimensions, 
depending on whether the application is 2D or 3D respectively. The updates from 
the tridiagonal solver, \texttt{Tridslv}s is accumulated to $u$ before the next 
iteration. For the 3D ADI application, there are three calls to 
\texttt{Tridslv}. An initial design implements it as a single hardware unit 
given the data dependencies between the calls. This enables FPGA resource 
utilization to be maximized by implementing 6 CUs each having 8 Thomas solvers 
synthesized as a vectorized solver. The \texttt{RHS} calculation, which is a 3D 
explicit stencil loop, was implemented following techniques in Kamalakkannan et 
al.~\cite{Kamalakkannan2021}, as a separate module. 
\vspace{-0.0cm}\begin{algorithm}
\begin{algorithmic}[1]
\FOR {$i = 0,i<n_{iter},i++$}
\STATE \texttt{Calculate RHS :} $d = f_{7pt}(u), a = \frac{-1}{2}\gamma, b = 
\gamma, c = \frac{-1}{2}\gamma $
\STATE \texttt{Tridslv(x-dim), update} $d$
\STATE \texttt{Tridslv(y-dim), update} $d$
\STATE \texttt{Tridslv(z-dim), update} $d$
\STATE $u = u + d$
\ENDFOR
\end{algorithmic}
\caption{\texttt{3D ADI Heat Application}}
\label{alg/adi}
\end{algorithm}\vspace{-0.0cm}The intermediate results 
between CUs and \texttt{RHS} module were written/read to/from external memory. 
The number of CUs is then limited by the available HBM ports but not by any 
other resource. An improvement on this initial design fuses the generation of 
$a,b,c$ coefficients with the tridiagonal solver. This enables the required 
number of HBM ports to be reduced and implementation of a maximum of 16 CUs. We 
opt for 12 CUs to reduce routing congestion which affects the maximum frequency 
achievable on the FPGA.

% \begin{figure*}[t]\centering
% \centering
% \subfloat[][DP - 100 iterations $v = 8, f_U = 3, N_{CU} = 6$ \textbf{update}]
% {\includegraphics[width=7.5cm]{Graphs/ADI-2D-DP}}\vspace{-5pt}
% \caption{\small 3D ADI Heat Diffusion Application 
% Performance}\vspace{-10pt}\label{grp/3DADI}
% \end{figure*}

The x-dim and y-dim solves can be synthesized as separate modules, pipelining 
the X and Y dimension calculation without needing to buffer intermediate 
results in external memory. Essentially, XY planes are buffered in on-chip 
memory to implement this, but it limits the mesh sizes solvable given 
the increased BRAM/URAM usage. To also pipeline the z-dim solve the full mesh 
must be buffered on-chip which significantly limits the mesh size, 
hence we do not attempt it here. The pipelining reduces the bandwidth 
requirement by half compared to the previous design. First module, 
\texttt{RHS+Tridslv(x-dim)+Tridslv(y-dim)} and second module 
\texttt{Tridslv(z-dim)} operate in parallel in a ping-pong fashion. This 
effectively increases the number of modules working in parallel to 24, 
considering the availability of HBM ports. The design now has a large pipeline 
start delay and is best utilized by batching large numbers of 3D meshes to 
obtain higher throughput. Xilinx data flow design synthesis requires separate 
data structures for independent read and write operations. We introduce two 
data structures for accumulation in line 6 of Algo \ref{alg/adi}. But due to 
limited HBM ports, we had to share single HBM ports with two data structures. 
This limits the data flow per data structure from/to the HBM2 ports as well as 
size of data structure, given single HBM port's capacity is 256MBs. This final 
design gave the best performance in our evaluations. The full pipeline latency 
for the design can be shown to be given by (\ref{eq/3dlats}): \vspace{-0.2cm}
\begin{equation}
L_{adi,3D} = n_{iter}\times MAX(L_{rhs+xy}, L_{z}) \label{eq/3dlats}
\end{equation}\vspace{-0.6cm}
\begin{align}
L_{rhs+xy} =& (\frac{2xy}{v}) + (2v\frac{x}{v} + 3gx) + 
(\frac{2xy}{v} + 3gy)+ \nonumber\\ & 
\ceil*{B/2N_{CU}}(\frac{xyz}{v})\label{eq/3dlats1}
\end{align}\vspace{-0.5cm}
\begin{equation}
L_{z} = (\frac{2xz}{v} + 3gz) + 
\ceil*{B/2N_{CU}}(\frac{xyz}{v})\label{eq/3dlats2}\vspace{-0.0cm}
\end{equation}
Here, $x,y$ and $z$ are the size of systems in each dimension, $N_{CU}$ is the 
number of CUs implemented on the FPGA and $B$ is the total number of 3D meshes, 
i.e the number of batches. The first term in (\ref{eq/3dlats1}) accounts for the 
latency of the 3D stencil computation in \texttt{RHS} which operates over 3 
planes. Here we read $xy$ number of mesh points in groups of the vectorization 
factor $v$. The second term is for \texttt{Tridslv(x-dim)} including latency to 
transpose the x-lines when reading/writing. Similarly the third term is for 
\texttt{Tridslv(y-dim)} including the read/write y-lines from the buffered 
x-lines. The final term is the latency to process $B$ number of meshes using 
$N_{CU}$ number of CUs. In (\ref{eq/3dlats2}) the first term is for the 
read/write and solving in the z dimension with \texttt{Tridslv(z-dim)}. We 
take the maximum in (\ref{eq/3dlats}) because the two modules need to be 
synchronized, as they are going to swap their read and write location after 
processing $B/2$ meshes. The vectorization factor $v$ is 8 for our design and 
$g$ is 32 for FP32 and 64 for FP64. A minor consideration for obtaining improved 
predictions from the above model is when the number of points per clock cycle 
arriving to the vectorized solvers is different to $v$ due to memory bandwidth. 
For example if we share a single HBM port to read two data structures and if we 
use 256 bit data path a lower number of points $p$ will come through the data 
path than $v$. Then, replacing $v$ by $p$ will be more accurate.

\begin{table}[t]\footnotesize
% \scriptsize
\centering \vspace{-0pt}
\caption{ \small ADI Heat Diffusion Application: Achieved Bandwidth, $BW$ 
(GB/s) 
and Energy, $E$ (J)}\vspace{-5pt}
\setlength\tabcolsep{0.2cm}
\begin{tabular}{@{}rrrrrrrrrrr@{}}
\toprule
\multicolumn{11}{c}{2D FP32 (120 iterations, $f_U = 3$), F - FPGA, G - 
GPU} \\
\midrule
Mesh  &  \multicolumn{3}{r}{$BW$-1500B} & \multicolumn{2}{r}{$E$-1500B}&   
\multicolumn{3}{r}{$BW$-3000B} & \multicolumn{2}{r}{$E$-3000B}\\
&  F & Gx &   Gy & F &  G & F &Gx & Gy & F & G\\
\midrule
$32^2$ 	& 501	&134		&418 	&1	&7		&563	&164	&493	&2	&13  \\
$48^2$ 	& 551	&168		&472	&2	&14		&596	&187	&507	&3	&26 \\
$64^2$ 	& 524	&184		&528 	&3	&23		&556	&199	&553	&6	&42 \\
$80^2$ 	& 597	&191	    &523 	&5	&39	    &621	&202	&533	&9	&72 \\
$96^2$ 	& 604	&201	    &541 	&7	&51	    &627	&207	&543	&13	&99 \\
$112^2$ & 602	&203	    &539 	&9	&71 	&626	&208	&535	&17	&138 \\
$128^2$ & 602	&207	    &563 	&12	&86 	&620	&206	&565	&23	&168 \\
\toprule
\end{tabular}\vspace{0pt}\label{tab/ADI_SP_BW_Energy}
\setlength\tabcolsep{0.2cm}
\begin{tabular}{@{}rrrrrrrrrrr@{}}
\multicolumn{11}{c}{2D FP64 (120 iterations, $f_U = 2$), F - FPGA, G - 
GPU} \\
\midrule
Mesh  &  \multicolumn{3}{r}{$BW$-1500B} & \multicolumn{2}{r}{$E$-1500B}&   
\multicolumn{3}{r}{$BW$-3000B} & \multicolumn{2}{r}{$E$-3000B}  \\
&  F & Gx &   Gy & F  &  G & F &Gx & Gy & F & G\\
\midrule
$32^2$	&360	&184    &508	&2	&10		&395	&196	&543	&4		&21 
\\
$48^2$	&377	&199    &527	&5	&24		&402	&206	&536	&10		&51 
\\
$64^2$	&380	&203    &557	&9	&42	    &399	&203	&529	&18		&88 
\\
$80^2$	&402	&209    &550	&14	&68	    &418	&209	&542	&26		&142 
\\
$96^2$	&408	&209    &557	&20	&98 	&421	&211	&563	&38		&199 
\\
$112^2$	&411	&208    &543	&26	&139	&424	&209	&549	&51		&277
\\
$128^2$	&411	&204    &517	&34	&179	&422	&210	&551	&67	    &355 
\\
\toprule
\end{tabular}\vspace{0pt}
\setlength\tabcolsep{0.1cm}
\begin{tabular}{@{}rrrrrrrrrrrrr@{}}
\multicolumn{11}{c}{3D FP32 (100 iterations), F - FPGA, G - 
GPU} \\
\midrule
Mesh  &  \multicolumn{4}{r}{$BW$-24B} & \multicolumn{2}{r}{$E$-24B}&   
\multicolumn{4}{r}{$BW$-72B} & \multicolumn{2}{r}{$E$-72B}  \\
&  F & Gx &   Gy &   Gz & F  &  G & F &Gx & Gy &   Gz& F & G\\
\midrule
$32\times32\times32$	&218	&119	&218	&288	&1	&4	    &266	&172		
&384 &493	&3	&9 \\
$80\times32\times32$	&252	&136	&364	&474	&2	&8	    &323	&188		
&411 &539	&5	&23 \\
$48\times48\times48$	&288	&171	&355	&475	&3	&11	    &338	&198		
&399 &551	&7	&31 \\
$80\times64\times64$	&326	&194	&412	&548	&7	&31	    &351	&207	 
&438 &561	&27	&70 \\
$80\times80\times80$	&337	&201	&417	&556	&11	&48	    &353	&203	
&394 &543	&31	&150 \\
$96\times96\times96$	&346	&210	&429	&568	&18	&78	    &358	&211	
&425 &563	&53	&241 \\
\toprule
\end{tabular}\vspace{-0.0cm}

\setlength\tabcolsep{0.1cm}
\begin{tabular}{@{}rrrrrrrrrrrrr@{}}
\multicolumn{11}{c}{3D FP64 (100 iterations), F - FPGA, G - 
GPU} \\
\midrule
Mesh  &  \multicolumn{4}{r}{$BW$-24B} & \multicolumn{2}{r}{$E$-24B}&   
\multicolumn{4}{r}{$BW$-72B} & \multicolumn{2}{r}{$E$-72B}  \\
&  F & Gx &   Gy &   Gz & F  &  G & F &Gx & Gy &   Gz& F & G\\
\midrule
$32\times32\times32$	&201	&165	&358	&445	&2	&6	    &239	&193		
&420 &527	&6	&17 \\
$80\times32\times32$	&222	&182	&406	&531	&5	&15	    &262	&204		
&419 &548	&14	&44 \\
$48\times48\times48$	&242	&194	&401	&536	&7	&20	    &267	&207		
&420 &554	&18	&59 \\
$80\times64\times64$	&262	&205	&427	&561	&18	&20	    &274	&209	
&426 &563	&52	&173 \\
$80\times80\times80$	&265	&209	&431	&564	&28	&90	    &271	&209	
&423 &563	&82	&275 \\
$96\times96\times96$	&271	&205	&426	&550	&47	&155	&276	&211	
&442 &565	&139 &464 \\
\toprule
\end{tabular}\vspace{-0pt}
\label{tab/ADI_DP_BW_Energy}\normalsize\vspace{-0.2cm}
\end{table}

A similar design can be developed for the 2D ADI application, but now the 
functions in the iterative loop \texttt{RHS}, \texttt{Tridslv(x-dim)} and 
\texttt{Tridslv(y-dim)} can all be pipelined. This makes it possible to unroll 
the iterative loop by some factor $f_{U}$. Note that the variable $u$ is 
incremented each iteration (line 6 of Algo.~\ref{alg/adi}) where the previous 
value of $u$ needs to be input at the end of each unrolled iteration to carry out 
this increment. However \texttt{RHS} of each iteration also consume $u$ and thus 
we use a delay-buffer (similar to ones used in 
StencilFlow~\cite{licht2020stencilflow}) implemented as an HBM FIFO to feed the 
previous values of $u$ to the increment stage noted in line 6. Unrolling 
iterative loop reduced the total number of data structures in external memory. 
Hence we are able to assign dedicated ports for each data structure which enable 
better data flow throughput. The performance model for the 2D application is 
given in (\ref{eq/2dlats}).\vspace{-0.0cm}
\begin{equation}
L_{adi,2D} = (n_{iter}/f_{U}) \times L_{rhs+xy} \label{eq/2dlats}
\end{equation}\vspace{-0.7cm}
\begin{align}
L_{rhs+xy}  =& f_{U}\times\left[ (\frac{2x}{v}) + (2v\frac{x}{v} + 3gx) + 
              (\frac{2xy}{v} + 3gy) \right] + \nonumber\\
             & \quad  \quad \ \ceil*{B/N_{CU}}\frac{xy}{v} 
\label{eq/2dlats1}\vspace{-0.4cm}
\end{align}
Pipeline latency increases with the unroll factor $f_{U}$, but for large 
$B$ it results in a higher overall speedup. The size of the FIFO delay 
buffer is equivalent to the total delay of \texttt{RHS}, 
\texttt{Tridslv(x-dim)} and \texttt{Tridslv(y-dim)} : $2x/v + 2vx/v+ 3gx + 
3gy + 2xy/v$.

% Results Analysis 2D ADI (FP32 and FP64)-
% * What is effective BW here ? what is the actual BW to memory (examples)
% * Batching configs really show the utility of the FPGA solution
% * Unrolling effectively doubles effective BW (?)
% * Model Predictions are very close with less than 10% error
% * FPGA booseted due to internal coefficient generation - in FPGA, we have 
% data movement with u,d, and acc1, acc2 in GPU, it is a,b,c,d,u, acc1, acc2 
% data structures
% * GPU x-solve have poor BW performance due to ... (transpose?)
% * Considering power draw the FPGA is even more superior 

% Results Analysis 3D ADI (FP32 and FP64)-
% How was batching done on the GPU ? How was it done on the FPGA 

% For the y-solve we used the \texttt{tridsolvlib} from from L\'{a}szl\'{o} et 
% al.~\cite{Laszlo2016}. However for the the x-solve we used the 
% \texttt{cusparse<t>gtsv2StridedBatch()} function from NVIDIA's cuSPARSE library
% as it provided marginally better performance for smaller meshes sizes and batch 
% sizes. 

% $$ 

\figurename{~\ref{grp/ADI}} details the performance of the 2D ADI Heat 
diffusion application implemented in both FP32 (a) and FP64 (b) on the Alveo 
U280 and compares it to execution on the V100 GPU. The design parameters for 
each are noted in the graphs. Operating frequencies are 292MHz and 288MHz for 
FP32 and FP64 respectively. In both cases the coefficients $a,b$ and $c$ are 
internally generated, on the FPGA. This means that only $u$ is read. Performance 
results demonstrate the FPGA outperforming the GPU particularly for runs with 
large batch sizes. Additionally the predictive model accuracy is over 85\% with 
large batched predictions being more accurate at over 90\%. Inspecting the 
effective bandwidth on each device as detailed in the top two sub-tables in 
\tablename{~\ref{tab/ADI_SP_BW_Energy}} provides insights into the superior 
performance of the FPGA. The bandwidth is computed by counting the total number 
of bytes transferred during the execution of each call in Alg.~\ref{alg/adi}, 
looking at the mesh data accessed and dividing it by the total time taken by 
each call. On the GPU, we have detailed the achieved bandwidth of the x- 
(Gx) and y-dim (Gy) solves. On the FPGA we show the full bandwidth achieved in 
the pipeline. The x-dim bandwidth on the GPU is significantly worse due to the 
$8\times8$  transpose operations. Such lower bandwidths are also confirmed by 
L\'{a}szl\'{o} et al.~\cite{Laszlo2016}. We additionally confirmed the same 
performance when using cuSPARSE's \texttt{cusparse<t>gtsv2StridedBatch()} 
library function for the x-solve. The higher performance of the FPGA can 
be attributed to the unrolling of the iterative loop allowing higher bandwidth 
utilization for the data path and the internal generation of coefficients. The 
GPU tridiagonal solver library is not currently setup for such an optimization. 
Thus, the application writes $a,b,c$ and $u$ to global memory after \texttt{RHS} 
and intermediate results also written/read between the two \texttt{Tridslv} 
calls whereas on the FPGA these stay on-chip. Even with modifications to the 
GPU library to generate coefficients internally which would improve GPU 
performance, we believe the FPGA results point to a very competitive solution, 
particularly when batching large meshes that can fit within the resource 
constraints of the FPGA implementation, for this application. 
\begin{figure*}[t]\centering
\centering
\subfloat[][\centering\footnotesize Thomas-PCR, $v=8, N_{CU} = 6$]
{\vspace{-0.2cm}\includegraphics[width=6.5cm]{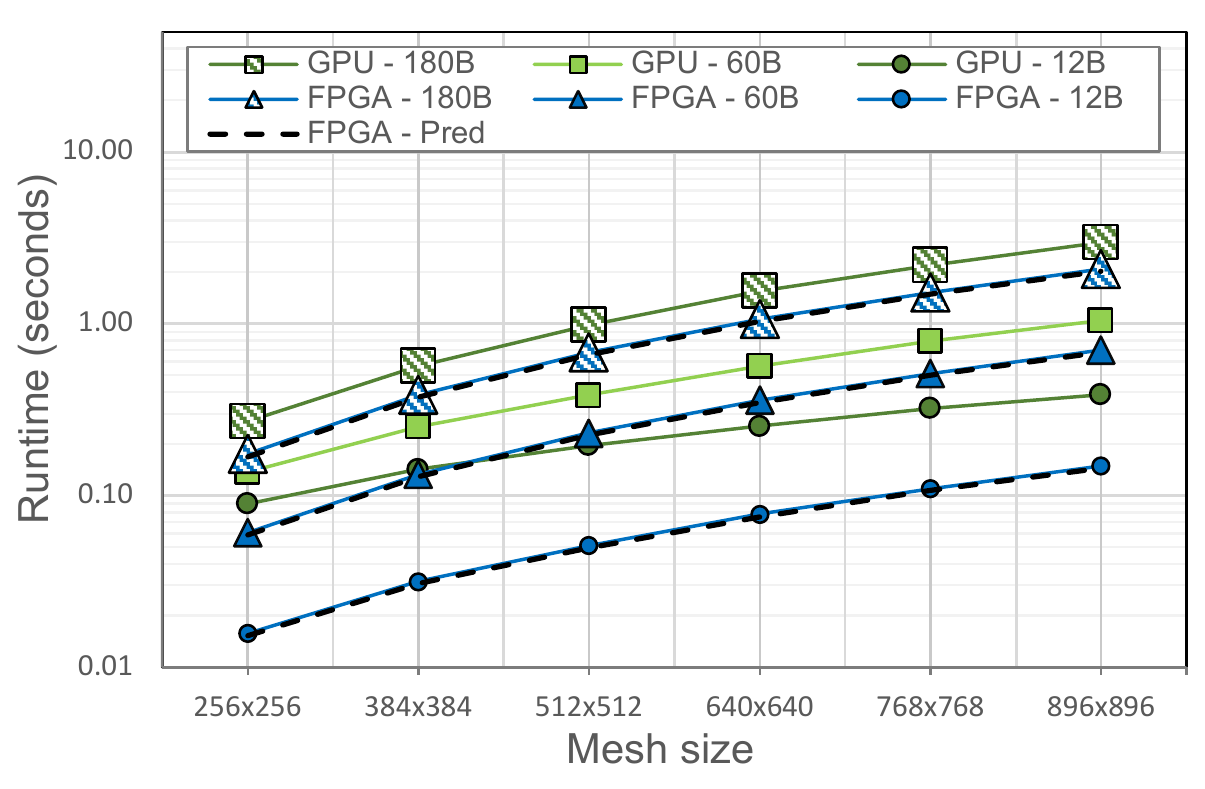}}
\hspace{
0.4cm}
\subfloat[][\centering\footnotesize Thomas-Thomas, $v=8, N_{CU} = 6$]
{\vspace{-0.2cm}\includegraphics[width=6.5cm]{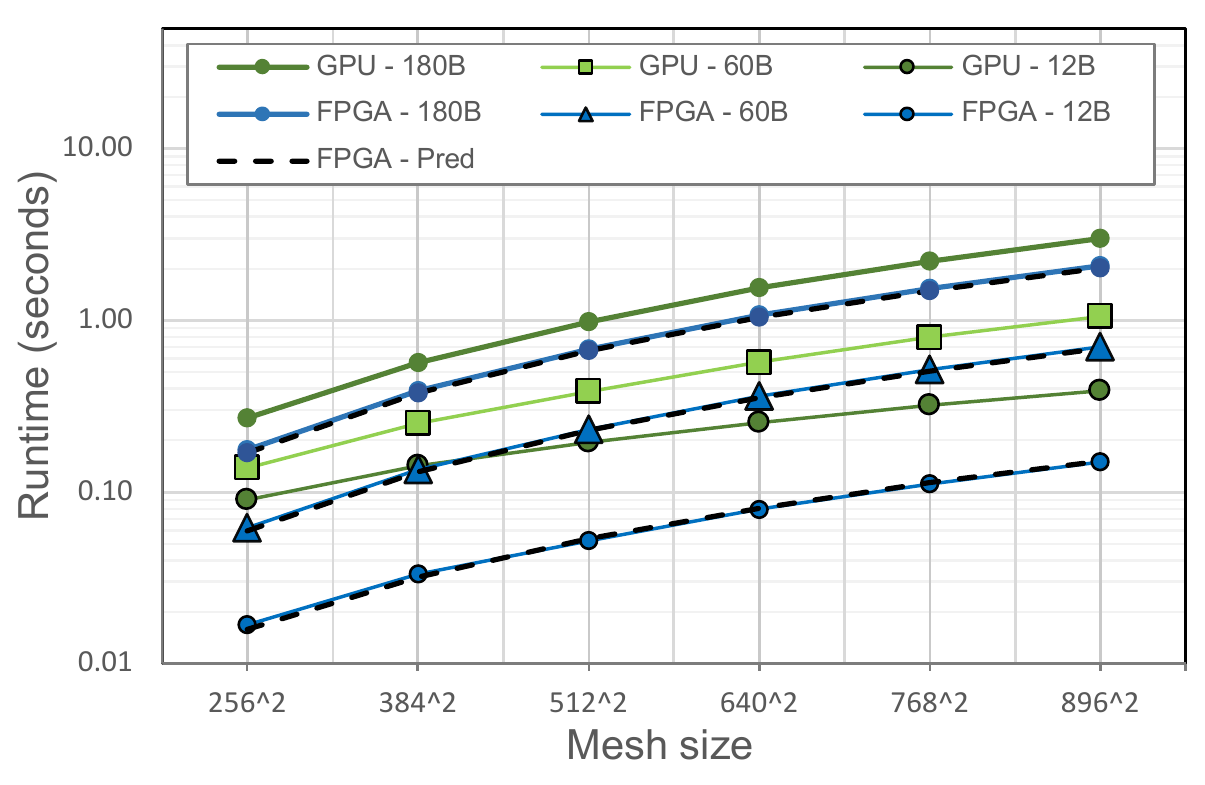}}
\hspace{
0.4cm}
\caption{\small 2D ADI large, FP32, 100 iter 
meshes}\vspace{-0.5cm}\label{grp/2DADItiled}
\end{figure*}

% \begin{figure}[t]\centering
% \centering
% \includegraphics[width=6.5cm]{Graphs/ADI-2D-SP_THPCR_log}\vspace{-0.3cm}\caption
% {\centering\small 2D ADI large meshes (Thomas-PCR, FP32, 100 
% iter)}\vspace{-0.5cm}\label{grp/2DADItiled_THPCR}\vspace{-0.2cm}
% \end{figure}

The first two sub-tables in~\tablename{~\ref{tab/ADI_SP_BW_Energy}} 
also detail the energy consumption of the 2D runs. The \texttt{xbutil} utility 
was used to measure power during FPGA execution, while \texttt{nvidia-smi} was 
used for the same on the V100. The FPGA on average consumed 75W while the GPU 
power draw ranged from 50W to 250W. Results indicate that the FPGA energy 
consumption is approximately $5\times$ to $6\times$ lower for this 2D 
problem. \figurename{~\ref{grp/ADI}}(c) and (d) and the bottom two sub-tables 
in \tablename{~\ref{tab/ADI_SP_BW_Energy}} detail the performance of the 3D ADI 
heat diffusion application in FP32 and FP64 respectively. Again we see 
performance trends similar to the 2D case, however we were only able to run 
smaller batch sizes due to HBM memory limitations for 3D meshes. On the GPU,
again apart from the x-dim solve we observe good achieved bandwidth. On the 
FPGA the achieved bandwidth is poorer due to no unrolling of the iterative loop 
as done in the 2D case, where there are 3CUs each unrolled by a factor of 3. The 
sharing of HBM ports as described in the design of this application limits the 
data flow per data structure further reducing achieved bandwidth. The energy 
consumption of the FPGA is $3$--$4\times$ less than on the GPU. 

A Thomas-Thomas based implementation for the 2D ADI-Heat application for larger 
mesh sizes can be modeled using (\ref{eq/2dthomasthomas}). 
% in \texttt{RHS} assume the tile sizes are $T_x$ and $T_y$. The model in 
% (\ref{eq/2dthomasthomas_rhs}) is obtained with techniques 
% in~\cite{Kamalakkannan2021}, accounting for the stencil order $D$ and 
% overlapping regions required to carry out redundant computations over the 
% border of tiles.
\vspace{-0.2cm}\begin{equation}
L_{adi,2D,tiled} = n_{iter}\times (L_{rhs+x} +L_{y})\label{eq/2dthomasthomas}\vspace{-0.3cm}
\end{equation}
\begin{align}
% L_{rhs  }  =& 2\frac{x}{v} 
% + By\ceil{\frac{y}{T_{y}-D}}\times 
% T_{y}\frac{x}{v} 
% \label{eq/2dthomasthomas_rhs}\\
L_{rhs+x}    =& 2\frac{x}{v} + 2v\frac{x}{v}+\frac{3gx}{t_{1}} + 4gt_{1}  + 
\frac{Bxy}{v}\label{eq/2dthomasthomasx}\\
L_{y}    =& 2y\frac{T_{x}}{v}+ \frac{3gy}{t_{2}} + 4gt_{2}+ 
\frac{Bxy}{v}\label{eq/2dthomasthomasy}
\end{align}\vspace{-0.3cm}
% $D$ determines $T_x$ and $T_y$.

% (note these are different to $T_x$ and $T_y$), 
\noindent In this case, \texttt{RHS} and x-solve can be pipelined but y-solve 
can't be pipe-lined together as we are computing ``tiles'' along the y-dim 
lines, huge internal memory will be required to transpose the mesh. 
The explicit stencil computation in RHS does not require tiling as we are 
not processing very large meshes. If the tile sizes for the Thomas-Thomas 
solvers are selected to be $t_1$ and $t_2$ then the reduced system sizes will be 
$2t_1$ and $2t_2$. Equation (\ref{eq/2dthomasthomasx}) accounts for the latency 
for \texttt{RHS} with x-dimension solve where the first term is stencil 
latency, second term is the latency for the data path, third is for the modified 
Thomas solve and fourth is the reduced solve. Similarly 
(\ref{eq/2dthomasthomasy}) given the y-dimension solve latency. Note that here 
we have used $T_x$ (this is different to $t1$) as the tile size for the x-dim 
data path where we buffer $T_{x}\times y$ sized planes. Note also that we have 
selected the number of interleaved systems and interleaved reduced systems to be 
equal (i.e. $g = g_r$ in relation to (\ref{eq/thomasthomas})). The final term 
in (\ref{eq/2dthomasthomasx}) and 
(\ref{eq/2dthomasthomasy}) simply gives the latency for processing a batch of B 
systems. Replacing the reduced system solve with the PCR algorithm is also 
possible where then the $4gt_{1}$ and $4gt_{2}$ terms in 
(\ref{eq/2dthomasthomasx}) and (\ref{eq/2dthomasthomasy}) will become 
$log(2t_{1})\times(2t_{1}+l)$ and $log(2t_{2})\times(2t_{2}+l)$.

% For Thoams-PCR
% 17 & 18 will get changed
% 4gt1 should be log(2t1)*(2t1+l)
% similarly 4gt2 will be

\begin{table}[t]\footnotesize
\scriptsize
\centering \vspace{-0pt}
\caption{ \small ADI Heat Diffusion Application on large meshes: Achieved 
Bandwidth, $BW$ (GB/s) and Energy, $E$ (J)}\vspace{-5pt}
\setlength\tabcolsep{0.10cm}
\begin{tabular}{@{}rrrrrrrrrrrrrrr@{}}
\multicolumn{15}{c}{ \tiny 2D FP64 (100 iter),  F1 - FPGA(Th-PCR), F2 - 
FPGA(Thomas-Thomas), G - GPU} \\
\midrule
Mesh  &  \multicolumn{4}{r}{$BW$-60B} & \multicolumn{3}{r}{$E$-60B}&   
\multicolumn{4}{r}{$BW$-180B} & \multicolumn{3}{r}{$E$-180B}  \\
&  F1 & F2 & Gx &   Gy & F1 & F2  &  G & F1 & F2 & Gx & Gy & F1 & F2 & G\\
\midrule
$256^2$		&206    &203	&117	&238	&5  &5 		&14		&217    &215	
&186		&464	&13     &13		&35 \\
$384^2$		&213    &209	&152	&323	&10	&10	    &29		&220    &218	
&204	    &530	&29	    &29	    &74 \\
$512^2$		&218    &217	&177	&400	&17 &17		&48		&222    &222	
&210	    &544	&51	    &51     &128 \\
$640^2$		&218    &217	&191	&450	&27 &27		&74		&221    &221	
&211	    &551	&80	    &80     &217 \\
$768^2$		&220    &219	&124	&471	&39 &39		&103	&223    &222	
&211	    &559	&114	&115    &308 \\
$896^2$		&220    &219	&204	&503	&53 &53	    &142	&222    &222	
&214	    &566	&156	&157    &418 \\
% $1024^2$	&222		&154	&596	&68	    &203	&223	&135	&302	
% &203	&523 \\
\toprule
\end{tabular}\vspace{-0pt}
\label{tab/ADI_Tiled_BW_Energy}\normalsize\vspace{-0.5cm}
\end{table}

\figurename{~\ref{grp/2DADItiled}} and 
\tablename{~\ref{tab/ADI_Tiled_BW_Energy}} presents the performance of 2D ADI 
heat diffusion application on large meshes solved using Thomas-PCR and 
Thomas-Thomas hybrid implementations. Again we compare with the same mesh sizes 
solved on the GPU. Due to the \texttt{RHS} and \texttt{Tridslv(x-dim)} being 
pipelined together, the FPGA gets better HBM bandwidth utilization. The GPU 
also gets good bandwidth utilization where it reaches bandwidth levels similar 
to batched smaller meshes. The FPGA can be seen to be $2\times$ to $3\times$ 
more energy efficient than the GPU for the largest mesh sizes.

\begin{figure*}[t]\centering
\centering
\subfloat[][\centering 40$\times$20 mesh, $v = 1, f_U = 1, N_{CU} = 3$]
{\includegraphics[width=6.5cm]{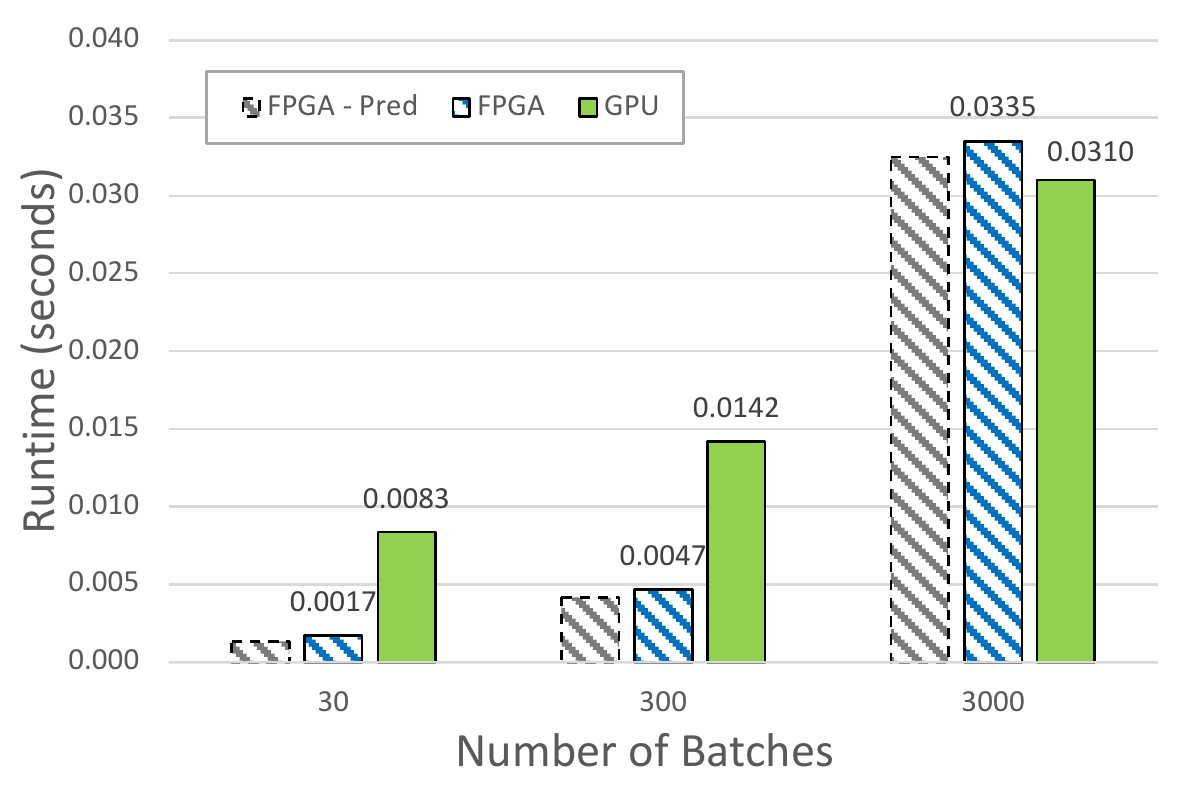}\vspace{-5pt}}\hspace{5pt}
\subfloat[][\centering 100$\times$50, $v = 1, f_U = 1, N_{CU} = 3$]
{\includegraphics[width=6.5cm]{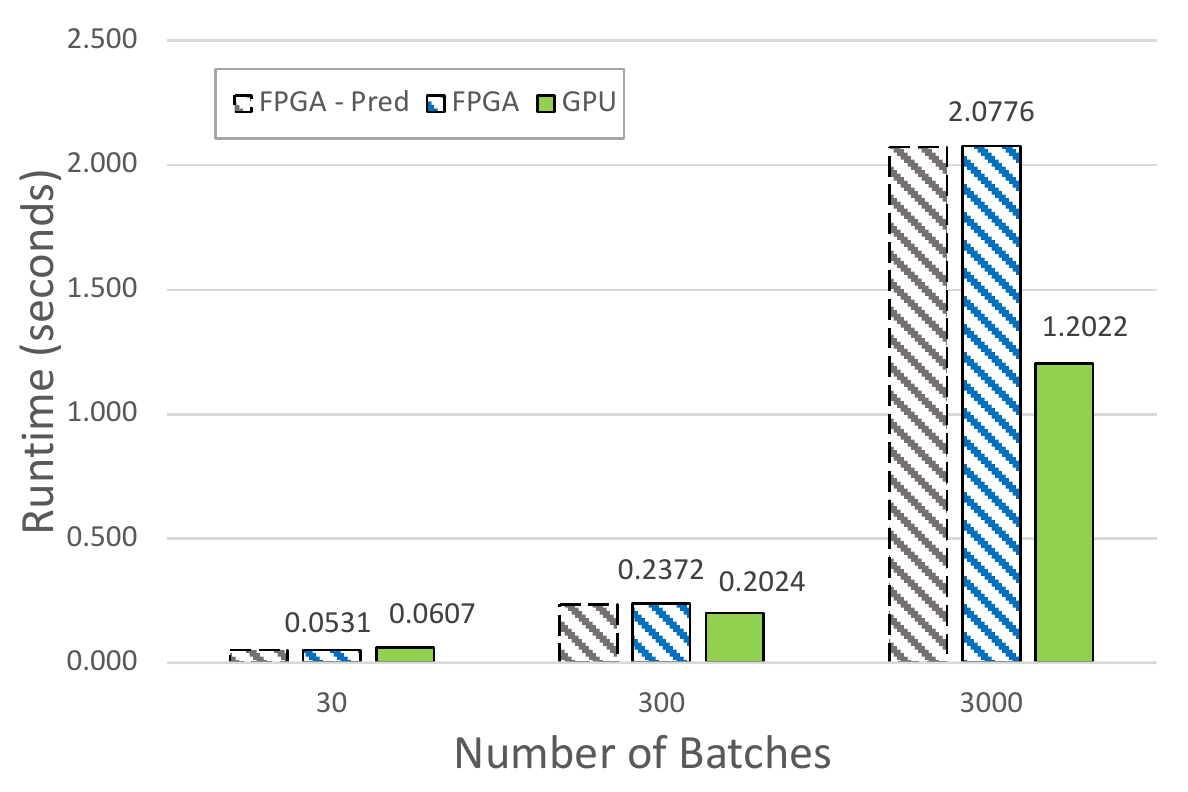}\vspace{-5pt}}\vspace{-5pt}
\caption{\small SLV application performance.}\vspace{-0.4cm}\label{grp/SLV}
\end{figure*}

\vspace{-0.2cm}
\subsection{SLV}
\vspace{-0.1cm}
\noindent The second application we synthesize and evaluate comes from
computational finance. It implements a stochastic local volatility (SLV) 
model, which describe asset price processes, particularly foreign exchange 
rates~\cite{Tataru2010}. A batched GPU implementation based on a 2nd order 
finite-difference scheme was developed for this problem using the OPS DSL 
by Reguly et al.~\cite{RegulyBatching2019}. It is a 2D application implemented 
in FP64 representation. Its high-level algorithm is detailed in 
Algo.~\ref{alg/slv}. \vspace{-0.3cm}\begin{algorithm}\begin{algorithmic}[1]
\FOR {$i = 0,i<n_{iter},i++$}
\STATE \texttt{hv\_pred0()}, \texttt{hv\_matrices()}
\STATE \texttt{Tridslv(x-dim)} 
\STATE \texttt{hv\_pred1(), Tridslv(y-dim)}
\STATE \texttt{hv\_pred2(), Tridslv(x-dim)}
\STATE \texttt{hv\_pred3(), Tridslv(y-dim)}
\ENDFOR
\end{algorithmic}
\caption{\texttt{2D Heston SLV Backward}}
\label{alg/slv}
\end{algorithm}\vspace{-0.35cm} The application implements a Hundsdorfer-Verwer 
(HV) method for time integration. The Rannacher smoothing available in the 
original application has been switched off in our evaluation. The 
\texttt{hv\_pred}* and \texttt{hv\_matrices} are explicit loops each using 10 
point stencils, requiring a window buffer 
implementation~\cite{Kamalakkannan2021} for data reuse. The 9 kernels in 
Algo.~\ref{alg/slv} were implemented as separate hardware modules pipelining the 
computation within the iterative loop. \texttt{hv\_matrices} generates a number 
of 2D coefficients \texttt{AX,BX,CX,AV,BV,CV} and 1D coefficient \texttt{EV} for 
the \texttt{Tridslv}s. Coefficients \texttt{AX,BX,CX} then needs to be input to 
(consumed by) \texttt{Tridslv(x-dim)} kernels and coefficients \texttt{AV,BV,CV} 
and \texttt{EV} to \texttt{Tridslv(y-dim)} kernels. Therefore these coefficients 
are consumed at different stages of the pipeline. However other inputs to the 
\texttt{Tridslv} calls come through the computation of this multi-stage 
pipeline. Therefore large FIFO delay buffers are required to keep 
synchronization (i.e. avoid pipeline stalling). As such we opt to regenerate 
the above coefficients at separate stages, essentially duplicating the 
circuitry. This results in the generation of coefficients \texttt{AX,BX,CX}, for 
the \texttt{Tridslv(x-dim)}, being fused to \texttt{hv\_pred0()} and 
\texttt{hv\_pred2()} and the generating of coefficients \texttt{AV,BV,CV,EV}, 
for the \texttt{Tridslv(y-dim)}, being fused to \texttt{hv\_pred1()} and 
\texttt{hv\_pred3()}. This results in a total of 8 hardware modules, requiring 
significantly smaller delay buffers than if we implemented the original set of 
kernels. The performance model for the SLV application is given in 
(\ref{eq/slvlats}):\vspace{-0.0cm}
\begin{align}
L_{slv} = &\  n_{iter} [ 4\times(2x)  + 2\times(3gx) + \nonumber\\
          & \quad \quad \ \ 2\times(3gy + 2xy) + \ceil*{B/N_{CU}}xy ] 
\label{eq/slvlats}\vspace{-0.3cm}
\end{align}
Here $g$ is 64 as SLV is in FP64. The first term is the combined input/output 
latency for the three explicit stencil computations in \texttt{hv\_pred*}. The 
second and third terms account for the calls for \texttt{Tridslv(x-dim)} 
and \texttt{Tridslv(y-dim)} calls respectively, including the read/write 
y-lines from the buffered x-lines. The final term is the latency for processing 
a batch size of $B$ 2D meshes through. The number of CUs, $N_{CU}$ for SLV on 
the FPGA was 3, given the considerably larger amount of DSP and memory resources 
required for the application, particularly due to its use of FP64 
representation. The FIFO delay-buffer size calculation was aided by the Xilinx 
HLS tools where the exact datapath pipeline latency was estimated to obtain 
buffer sizes adequate for an implementation. \\
\indent The motivation for batched solves of multi-dimensional tridiagonal 
systems primarily comes from financial computing where for example computing 
prices of financial options and managing of risk by hedging options leads to 
the need to solve Algo.~\ref{alg/slv} type applications with different 
sets 
of coefficients~\cite{RegulyBatching2019}. Additionally carrying out extensive 
speculative scenarios required by regulators under various market conditions to 
evaluate a bank's exposure means that there are large number of options in the 
order of thousands to hundreds of thousands to be computed every day. Such a 
workloads would entail large numbers of roughly identical PDE problems to be 
solved which are well suited to be batched together.

\begin{table}[t]\footnotesize
\centering \vspace{-0pt}
\caption{ \small SLV : Bandwidth, $BW$ (GB/s) and Energy, $E$, (J)}\vspace{-5pt}
\setlength\tabcolsep{5pt}
\begin{tabular}{@{}lrrrrrr@{}}
\midrule
\multicolumn{7}{c}{40$\times$20 mesh} \\
Batch  &  \multicolumn{3}{c}{$BW$} &\phantom{x}& \multicolumn{2}{c}{$E$} \\   
 \cmidrule{2-4}\cmidrule{6-7}
       &  F & Gx & Gy           &&  F & G             \\
\midrule
% 3	  &6.67	  &0.31	    &3.31	    &&0.11	&0.41 \\
30	  &55.24  &3.04	    &28.01      &&0.13	&0.45 \\
300	  &202.31 &16.48    &176.51     &&0.35	&1.02 \\
3000  &281.06 &123.84   &327.65     &&2.51	&4.75 \\
% 4000  &323.52 &183.39 &411.88 &&3.56	&6.18 \\
\midrule
\multicolumn{7}{c}{100$\times$50 mesh} \\
Batch  &  \multicolumn{3}{c}{$BW$} &\phantom{x}& \multicolumn{2}{c}{$E$} \\  
\cmidrule{2-4}\cmidrule{6-7}
        &  F & Gx & Gy           &&  F & G             \\
\midrule
% 3	  &19.04	    &7.08	 &16.80	  &&2.61	 &2.37 \\
30	  &124.63       &51.28   &109.65  &&3.98	 &3.76 \\
300	  &278.87       &235.22  &238.34  &&17.79	 &22.26 \\
3000  &318.36       &421.77  &429.21  &&155.82   &216.40 \\
\toprule
\end{tabular}\vspace{0pt}
\label{tab/SLV_BW_Energy}\normalsize\vspace{-0.5cm}
\end{table}

\figurename{~\ref{grp/SLV}} and \tablename{~\ref{tab/SLV_BW_Energy}} 
details the runtime, bandwidth and energy performance of the SLV 
application implementation. Only two specific mesh sizes were available from 
the authors of the original code~\cite{RegulyBatching2019} each was batched up 
to 3000 batches of 2D meshes for this evaluation. The application is 
significantly more complex given the additional explicit stencil loops as well 
as the tridiagonal solvers. The runtimes here were obtained with the FPGA 
operating at 253.5MHz. As can be seen from the figures, the FPGA in some cases 
is faster than the V100 GPU, but for the largest batch sizes we attempted here, 
it is 8\%-70\% slower than the GPU. However the FPGA solution is over 30\% more 
energy efficient for large batch solves over the GPU. The achieved bandwidth on 
the FPGA is approximately at the same level as the 2D ADI 64FP version. Runtime 
predictions from the model were also observed to be over 90\% accurate for all 
cases. 

Finally, the resource utilization for all the synthesized designs on the Alveo 
U280 is shown in \tablename{~\ref{tab/resources}}. Different mesh sizes are 
supported without re-implementing the FPGA design. For ADI with small mesh sizes,  
URAM availability is the limiting factor. Thomas-PCR is marginally limited by 
DSP units and BRAM and did not allow for a more resource intensive design due to 
routing congestion. The SLV application is very much limited by the available 
number of DSP units on the FPGA. The implementation has 3CUs where a single SLR 
unit on the U280 was required for a single CU.

\begin{table}[t]\footnotesize
%\footnotesize
\scriptsize
\centering \vspace{-0pt}
\caption{ \centering\small U280 resource consumption for synthesized 
designs}\vspace{-3pt}
\setlength\tabcolsep{0.05cm}
\begin{tabular}{@{}lrrrrrr@{}}
Application  &  \%LUT & \%Reg & \%DSP & \%BRAM &  \%URAM & \#HBM Ports\\
\midrule
2D ADI F32 		  &45.9 &39.0 &48.2 &37.1 &75.0 &24 \\
2D ADI F64 		  &42.8 &30.8 &35.2 &61.9 &85.0 &18 \\
3D ADI F32 		  &52.7 &42.9 &53.2 &35.4 &80.0 &24 \\
3D ADI F64 		  &32.9 &24.4 &27.3 &17.7 &75.0 &24 \\
Thomas-Thomas F32 &35.8 &31.1 &38.0 &41.7 &48.8 &24 \\
Thomas-PCR F32 	  &54.1 &40.1 &52.9 &42.7 &39.4 &24 \\
SLV               &48.5 &38.7 &64.6 &37.9 &52.2 &15 \\
\toprule
\end{tabular}
\label{tab/resources}\normalsize\vspace{-0.7cm}
\end{table}

\vspace{-0.3cm}
\section{Conclusion}\label{sec/conclusions}\vspace{-0.0cm}
\noindent We presented a design space exploration for synthesizing optimized 
high throughput multi-dimension tridiagonal systems solvers on FPGAs. The main 
algorithms for direct solution of multiple tridiagonal systems were evaluated 
in light of the significant parallelization opportunities afforded by this 
class of solvers, particularly exploitable through the data-flow programming 
model on FPGAs. We developed a new tridiagonal solver library aimed at 
implementing high-performance computing applications on Xilinx FPGAs. Key new 
features of the library include data-flow techniques and optimizations for 
gaining high throughput with batching multiple system solves, replication of 
circuitry to carry out multiple solves in a ``vectorized'' manner and 
utilization of HBM memory available on modern FPGAs. The best algorithm for the 
FPGA with batched systems proved to be the Thomas algorithm, even with its loop 
carried dependencies, due to its simplicity and lower resource consumption. This 
somewhat subverts the conventional expectation of the more parallel PCR or 
spike algorithms being the best suited to get higher performance on parallel 
architectures. Our library, compared to the current state-of-the-art Xilinx 
tridiagonal library based on the PCR algorithm provided further evidence, where 
we see the superior performance of our Thomas based solver for larger batch 
sizes. However, for larger mesh sizes a hybrid Thomas-PCR or Thomas-Thomas 
solution was required due to the limitations of on-chip memory and demonstrated 
good performance overall with batched configurations.\\
\indent Two representative applications (1) a heat diffusion problem based on 
the ADI method and (2) a stochastic local volatility (SLV) model from the 
financial computing domain that rely on the solution of  multi-dimensional 
tridiagonal systems were implemented using the new library on a Xilinx Alveo 
U280 FPGA. As part of the design process a predictive analytic model that 
estimates the runtime performance of FPGA designs was also developed to assist 
in design space evaluations. The performance achieved by the FPGA was compared 
to optimized solutions of the same applications on a modern Nvidia Tesla V100 
GPU, showing competitive performance, sometimes even surpassing that of the 
GPU. This was true for both small and larger mesh problems which enabled 
creating of longer pipelines keeping intermediate results on FPGA on-chip 
memory. \\
\indent Even when runtime is inferior to the GPU, significant energy savings, 
over 30\% for the most complex application (SLV) with large batch sizes, were 
observed. Considering the motivating real-world scenario for such an 
application from the financial computing domain, such energy savings point to a 
significant cost benefit in overall operation. The predictive model provides 
over 85\% accuracy illustrating its significant utility in developing 
profitable FPGA designs. \\
\indent The FPGA library, the 2D/3D ADI heat diffusion application and 
optimized GPU source code developed in this paper are available as open source 
software at~\cite{fpgatridslv}. This code repository also contains results from 
a Xilinx Alveo U50 FPGA, which was also done as part of this research to 
investigate the performance and portability of our multi-dimensional 
tridiagonal solver library. The U50 results also confirms the insights and 
conclusions from this paper. Future work will explore the use of FPGA hardware 
from Intel, currently the other major FPGA device vendor, for this class of 
applications. 

% These algorithms usually provide superior performance on traditional CPU or GPU 
% architectures. 

% Vectorization factor need to be noted, if not possible to vectorize, need to 
% give estimates of resources when doing this 
% Source node delays ?

% Detail application implementations + optimizations [ADI and SLV]\\
% Detail performance model for each application \\

% ADI - 2D,3D (FP32 and FP64) / GPU vs FPGA\\
% SLV - 2D only (FP64) / GPU vs FPGA\\
% Need to do runtime (including prediction) and power performance\\

% For each of the above we need to do Application specific optimizations : 
% * Small/Tiny meshes - targeting the use of on chip memory to get more 
% performance\\
% * Pipeline preproc, X-dim and Y-dim\\
% * Unroll iterative loop \\

% For FP64 , FADD , FMULT and FDIV need more latency.
% This computation comes thorough longer circuit, hence needs more register 
% stages in the middle to improve the clock frequency but this will increase 
% the required number of clock cycles
% it is specific to U280 class of Xilinx FPGAs
% for intel FPGAs, since they have hardware Floating point cores,
% FADD and FMUl latencies are lower

%While GPUs were still best for solving larger mesh problems the FPGA did come 
% close to matching performance. 

% \noindent\textbf{TODO ?? - comparison text to cuSPARCE}\\
% \noindent\textbf{TODO ?? - emphasize in text for model and in performance that 
% Xilinx lib is good for lower batch numbers}
% \textbf{TODO - Some important texts from the Rubuttal should be included}
% \textbf{TODO - link to U50 results}

\vspace{-0.4cm}
\section*{Acknowledgment} %\small
\noindent Gihan Mudalige was supported by the Royal Society Industry Fellowship 
Scheme (INF/R1/1800 12). Istv\'an Reguly was supported by National Research, 
Development and  Innovation Fund of Hungary (PD 124905), under the PD\_17 
funding scheme. We are grateful to Xilinx for their hardware and software 
donations and Jacques Du Toit and Tim Schmielau at NAG UK for their advice and 
making the SLV application avaialble for this work.\normalsize\vspace{-0.3cm}

\bibliographystyle{IEEEtranS}
\bibliography{ref}

\end{document}